\documentclass[aps,twocolumn,prc,superscriptaddress,showpacs,nofootinbib,floatfix,amssymb,amsfonts,amsmath]{revtex4-1}

\usepackage{graphicx}
\usepackage{dcolumn}
\usepackage{bm}

\usepackage{amsmath}    
\usepackage{amsfonts}   
\usepackage{amssymb}
\usepackage{graphicx}   
\usepackage{longtable}

\begin{document}
\title{Octupole deformation in the ground states of even-even
       $Z\sim 96, N\sim 196$ actinides and superheavy nuclei}

\author{S.\ E.\ Agbemava}
\affiliation{Department of Physics and Astronomy, Mississippi
State University, MS 39762}

\author{A.\ V.\ Afanasjev}
\affiliation{Department of Physics and Astronomy, Mississippi
State University, MS 39762}

\date{\today}

\begin{abstract}
   
     A systematic search for axial octupole deformation in the actinides 
and superheavy nuclei with proton numbers $Z=88-126$ and neutron numbers from two-proton drip line 
up to $N=210$ has been performed in covariant density functional theory (DFT) using 
four state-of-the-art covariant energy density functionals representing different 
model classes. The nuclei in the $Z\sim 96, N\sim 196$ region of octupole deformation 
have been investigated in detail and the systematic uncertainties in the description 
of their observables have been quantified. Similar region of octupole deformation 
exists also in Skyrme DFT and microscopic+macroscopic approach but it is centered 
at somewhat different particle numbers. Theoretical uncertainties in the predictions 
of the regions of octupole deformation are increasing on going to superheavy nuclei 
with $Z\sim 120, N\sim 190$. There are no octupole deformed nuclei for $Z=112-126$ 
in covariant DFT calculations. This agrees with Skyrme DFT calculations, 
but disagrees with Gogny DFT and microscopic+macroscopic calculations which predict 
extended $Z\sim 120, N\sim 190$ region of octupole deformation.  

\end{abstract}


\maketitle

\section{Introduction}

  Reflection asymmetric (or octupole deformed) shapes represent an
interesting example of symmetry breaking of the nuclear mean field
\cite{BN.96}.  They are present in the ground and rotating states of 
the lanthanides with $Z\sim 58, N\sim 90$ and light actinides with 
$Z\sim 90, N\sim 136$ (see Refs.\ \cite{BN.96,MBCOISI.08,RR.12,AAR.16} 
and references quoted therein). These shapes also affect the outer fission 
barriers in actinides and superheavy nuclei \cite{AAR.12,SR.16,Zhou.16} 
and cluster radioactivity \cite{WR.11}. The first significant wave 
of the studies of octupole deformed shapes took place in the 80ies and 
first half of 90ies of the last century (see review of Ref.\ \cite{BN.96}). 
The interest to the study of such shapes has significantly increased 
during this decade (see references on theoretical and experimental 
works quoted in Ref.\ \cite{AAR.16}).
 
 Different theoretical frameworks have been used for the study
of octupole deformed shapes (see Refs. \cite{BN.96,AAR.16,SR.16} 
and references quoted therein). Here we employ the covariant density 
functional theory (CDFT)~\cite{VALR.05}. Its previous applications
to the investigation of such shapes have been overviewed and 
compared with the results of non-relativistic studies in Ref.\ 
\cite{AAR.16}. Built on Lorentz covariance and the Dirac equation, 
CDFT provides a natural incorporation of spin degrees of freedom 
\cite{Rei.89,Ring1996_PPNP37-193} and a good parameter free 
description of spin-orbit splittings \cite{Ring1996_PPNP37-193,BRRMG.99,LA.11}, 
which have an essential influence on the underlying shell structure. 
In addition, in CDFT the time-odd components of the mean fields are 
given by the spatial components of the Lorentz vectors. Therefore, 
because of Lorentz invariance, these fields are coupled with the same 
constants as the time-like components \cite{AA.10} which are fitted 
to ground state properties of finite nuclei (which are affected
only by time-even mean fields) and nuclear matter properties.

 Starting from pioneering work of Ref.\ \cite{RMRG.95}, the CDFT has 
been extensively used in the study of reflection asymmetric shapes especially 
during last decade. Most of these applications have been focused on reflection 
symmetric shapes with axial symmetry; they have been reviewed in Ref.\ 
\cite{AAR.16}. Let us mention some of these studies performed in the actinides. 
At the mean field level, the ground state properties of the actinides have 
been studied in Refs.\ \cite{RMRG.95,GMT.07,LSYVM.13,NVL.13,NVNL.14,AAR.16}.
Some axial octupole deformed nuclei have been studied also in the beyond mean
field approaches based on CDFT. For example, simultaneous quadrupole and 
octupole shape phase transitions in the Th isotopes have been studied in Ref.\ 
\cite{LSYVM.13} employing microscopic collective Hamiltonian. Using Interacting 
Boson Model Hamiltonian with parameters determined by mapping the microscopic 
potential energy surfaces, obtained in the relativistic
Hartree-Bogoliubov calculations, to the expectation value of the Hamiltonian
in the boson condensate the microscopic analysis of the octupole phase 
transition has been performed in Refs.\ \cite{NVL.13,NVNL.14}. The generator 
coordinate method studies taking into account dynamical correlations and 
quadrupole-octupole shape fluctuations have been undertaken in $^{224}$Ra 
employing the PC-PK1 functional in Ref.\ \cite{YZL.15}. They revealed 
rotation-induced octupole shape stabilization.

 Nonaxial-octupole ${Y}_{32}$ correlations in the $N=150$ isotones and 
tetrahedral shapes in neutron-rich Zr isotopes have been studied in 
Refs.\ \cite{ZLEZ.12,ZLZZ.17} employing  multidimensional constrained 
CDFT. Although the energy gain due to $\beta_{32}$ distortion exceeds
300 keV in $^{248}$Cf and $^{250}$Fm in model calculations, it is not 
likely that static deformation of this type is present in nature in
these two nuclei.  This is because their rotational features are well 
described  in the cranked relativistic Hartree-Bogoliubov framework 
with no octupole deformation \cite{AO.13,A.14}. Despite theoretical 
predictions and substantial experimental efforts a clear  experimental 
signal for tetrahedral shapes is still absent (see the discussion in
the introduction of Ref.\ \cite{ZLZZ.17}). In addition, symmetry unrestricted 
multidimensional constrained CDFT calculations are extremely time-consuming.
Because of these reasons only reflection symmetric shapes with axial 
symmetry are considered in the present paper.

 The most comprehensive study of octupole deformed shapes at the
mean field level within the CDFT framework has been performed in Ref.\ 
\cite{AAR.16}. In this 
manuscript the global search for such shapes has been carried 
out in all $Z\leq 106$ even-even nuclei located between two-proton 
and two-neutron drip lines with two covariant energy density functionals 
(CEDFs) NL3* and DD-PC1. As a result, a new region of octupole deformation, 
centered around $Z\sim 98, N\sim  196$ has been found in the CDFT framework
for the first time. Based on the results obtained with these two functionals 
it was concluded that in terms of its size in the $(Z,N)$ plane and the impact 
of octupole deformation on binding energies this region is similar to the best 
known region of octupole deformed nuclei centered at $Z\sim 90, N\sim 136$. 
In addition, the systematic uncertainties in the description of the ground 
states of octupole deformed nuclei in the $Z\sim 58, N\sim 90$ lanthanides and 
$Z\sim 90, N\sim 136$ actinides have been defined for the first time 
in the CDFT framework using five state-of-the-art CEDFs representing 
different classes of the CDFT models.

  However, the number of questions still remains unresolved in Ref.\ 
\cite{AAR.16}. The search for the answers on these questions is the
main goal of this manuscript. First, there are the indications 
that octupole deformation can be present in the ground states
of superheavy elements (SHE) with $Z\geq 108, N\sim 190$. They come 
from the results of the calculations within the microscopic+macroscopic 
(mic+mac)
approach (Ref.\ \cite{MNMS.95}) and non-relativistic Hartree-Fock-Bogoliubov 
(HFB) method based on finite range Gogny D1S force (Ref.\ \cite{WE.12}). 
To our knowledge no search of octupole deformation in the ground states of 
superheavy $Z\geq 108$ nuclei has been performed within the CDFT framework 
so far. To fill this gap in our knowledge we will perform such a search in 
the region of proton numbers $108 \leq Z \leq 126$ and in the region of 
neutron numbers from the two-proton drip line up to neutron number $N = 210$. 
This region almost coincides with the region used in recent reexamination 
of the properties of SHE in the CDFT framework in Ref.\ \cite{AANR.15}.

  Second, we will establish systematic theoretical uncertainties in the 
predictions of the properties of the octupole deformed nuclei in the $Z\sim 98, 
N\sim  196$ mass region and in superheavy nuclei. This is important since 
these nuclei will not be accessible with future facilities such as FRIB. However, 
the accounting of octupole deformation in the ground states of these nuclei is 
essential for the modeling of fission recycling in neutron 
star mergers \cite{GBJ.11,JBPAGJ.15} since the gain in binding energy of 
the ground states due to octupole deformation will increase the fission barrier 
heights as compared with the case when octupole deformation is neglected.  

  To achieve these goals we use the four most up-to-date covariant energy density 
functionals of different types, with a nonlinear meson coupling (NL3* \cite{NL3*}), 
with density-dependent meson couplings (DD-ME2 \cite{DD-ME2}), and with 
density-dependent zero-range interactions (DD-PC1 \cite{DD-PC1} and PC-PK1 
\cite{PC-PK1}). They represent different classes of CDFT models (see discussion in 
Ref.\ \cite{AARR.14}). The functional DD-ME$\delta$ used in our previous studies 
of the global performance of CDFT \cite{AARR.13,AARR.14,AANR.15,AAR.16,AA.16,AAR.17} 
is not employed here since it fails to reproduce octupole deformation in light 
actinides \cite{AAR.16} and inner fission barriers in superheavy nuclei \cite{AAR.17}.

  The paper is organized as follows. Section \ref{theory_details} describes 
the details of the solutions of the relativistic Hartree-Bogoliubov equations.  
Sec.\ \ref{oct-def-dep} is devoted to the discussion of the ground state properties 
of octupole deformed nuclei and their dependence on the covariant energy density 
functional. The evolution of potential energy surfaces with proton and neutron 
numbers is discussed in Sec.\ \ref{PES-evolution}. The assesment of systematic 
theoretical uncertainties in the predictions of ground state properties of 
octupole deformed nuclei and the comparison with other model predictions are 
performed in Sec.\ \ref{uncertainties}. Finally, Sec.\ \ref{conclusions} 
summarizes the results of our work.

\section{The details of the theoretical calculations}
\label{theory_details}

 The calculations have been performed in the
Relativistic-Hartree-Bogoliubov (RHB) approach using
parallel computer code RHB-OCT developed in Ref.\ \cite{AAR.16}.
Note that only axial reflection asymmetric shapes are considered 
in this code. 

  The calculations in the RHB-OCT code perform the variation of the function
\begin{eqnarray}
E_{RHB} + \sum_{\lambda=2,3} C_{\lambda 0}
(\langle\hat{Q}_{\lambda 0}\rangle-q_{\lambda 0})^2
\label{constr}
\end{eqnarray}
employing the method of quadratic constraints. Here $E_{RHB}$ is the
total energy (see Ref.\ \cite{AARR.14} for more details of its
definition) and $\langle\hat{Q}_{\lambda 0}\rangle$ denote the expectation
value of the quadrupole ($\hat{Q}_{20}$) and octupole ($\hat{Q}_{30}$)
moments which are defined as
\begin{eqnarray}
\hat{Q}_{20}&=&2z^2-x^2-y^2,\\
\hat{Q}_{30}&=&z(2z^2-3x^2-3y^2).
\end{eqnarray}
$C_{20}$ and  $C_{30}$ in Eq.\ (\ref{constr}) are corresponding stiffness
constants \cite{RS.80} and $q_{20}$ and $q_{30}$ are constrained values of the
quadrupole and octupole moments. In order to provide the convergence to the
exact value of the desired multipole moment we use the method suggested in
Ref.~\cite{BFH.05}. Here the quantity $q_{\lambda 0}$ is replaced by the parameter
$q_{\lambda 0}^{eff}$, which is automatically modified during the iteration in such
a way that we obtain $\langle\hat{Q}_{\lambda 0}\rangle = q_{\lambda 0}$ for the
converged solution. This method works well in our constrained  calculations.
We also fix the (average) center-of-mass of the nucleus at the origin with
the constraint
\begin{eqnarray}
<\hat{Q}_{10}>=0
\end{eqnarray}
on the center-of-mass operator $\hat{Q}_{10}$ in order to avoid
a spurious motion of the center-of-mass.

\begin{figure}
\includegraphics[angle=0,width=8.8cm]{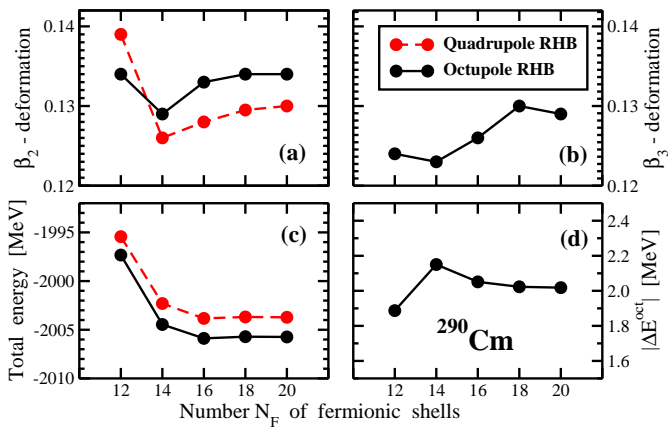}
\caption{(Color online) The dependence of calculated quadrupole 
and octupole deformations, total binding energy and the 
$|\Delta E^{oct}|$ quantity on the number of fermionic shells
employed in the RHB calculations for $^{290}$Cm with the DD-PC1 
functional. The results obtained in octupole and quadrupole RHB 
codes in respective local minima with $(\beta_2 \neq 0, \beta_3 \neq 0)$  
and $(\beta_2 \neq 0, \beta_3 = 0)$  are shown by solid black and 
dashed red curves, respectively. }
\label{truncation}
\end{figure}

  The charge quadrupole and octupole moments are defined as
\begin{eqnarray}
Q_{20} &=& \int d^3r \rho({\bm r})\,(2z^2-r^2_\perp), \\
Q_{30} &=& \int d^3r \rho({\bm r})\,z(2z^2-3r^2_\perp)
\end{eqnarray}
with $r^2_\perp=x^2+y^2$. In principle these values can be directly
compared with experimental data. However, it is more convenient to
transform these quantities into dimensionless deformation
parameters $\beta_2$ and $\beta_3$ using the relations
\begin{eqnarray}
Q_{20}&=&\sqrt{\frac{16\pi}{5}} \frac{3}{4\pi} Z R_0^2 \beta_2,
\label{beta2_def} \\
Q_{30}&=&\sqrt{\frac{16\pi}{7}}\frac{3}{4\pi} Z R_0^3 \beta_3
\label{beta4_def}
\end{eqnarray}
where $R_0=1.2 A^{1/3}$. These deformation parameters are more
frequently used in experimental works than quadrupole and octupole
moments. In addition, the potential energy surfaces (PES) are plotted 
in this manuscript in the ($\beta_2,\beta_3$) deformation plane.

\begin{figure*}
\includegraphics[angle=-90,width=8.8cm]{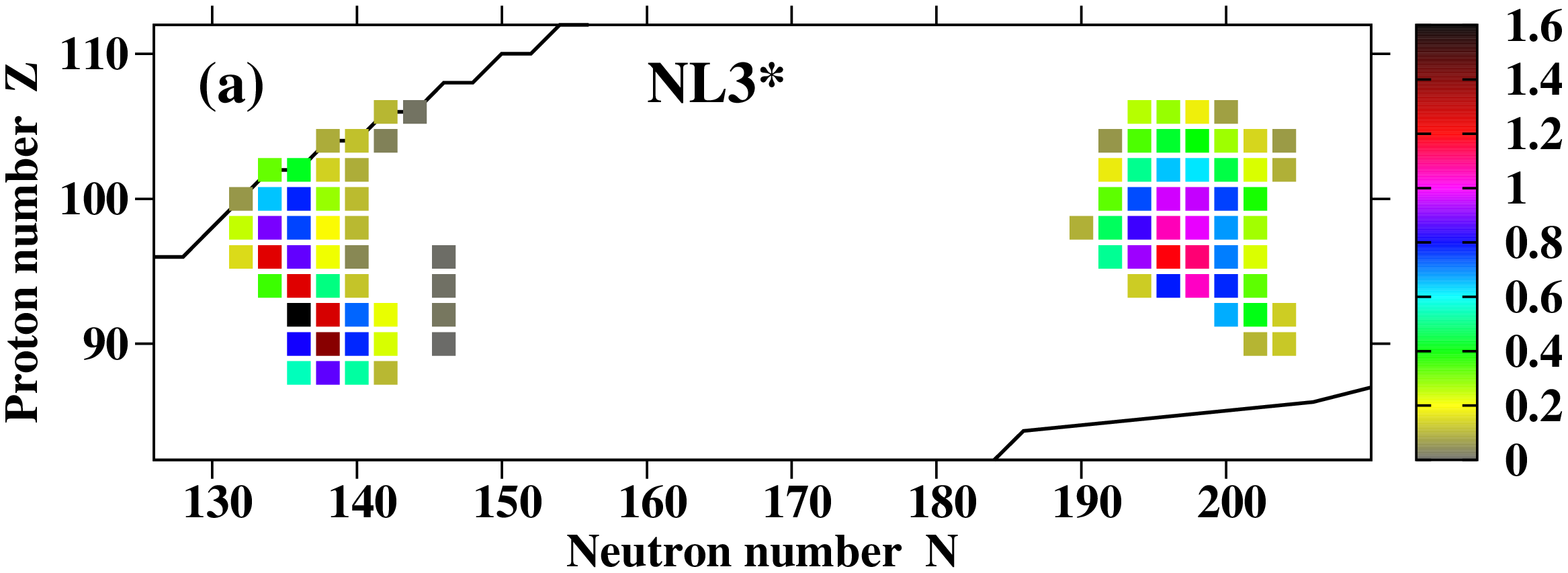}
\includegraphics[angle=-90,width=8.8cm]{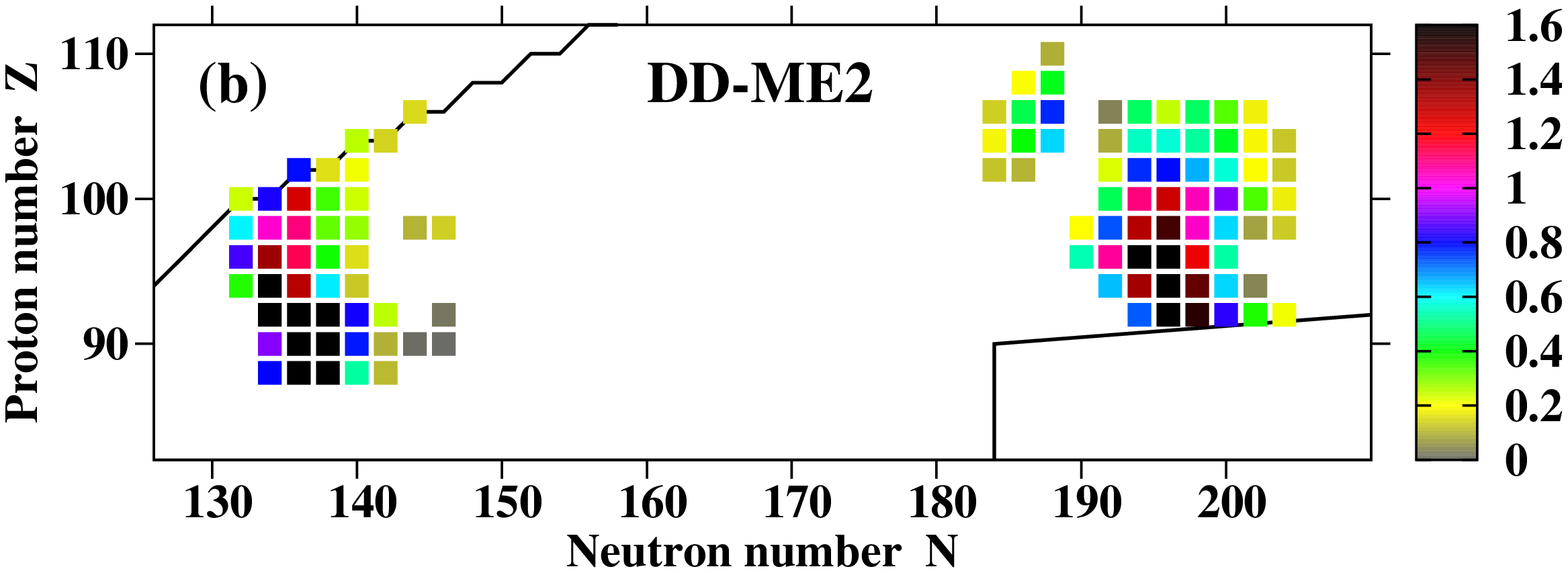}
\includegraphics[angle=-90,width=8.8cm]{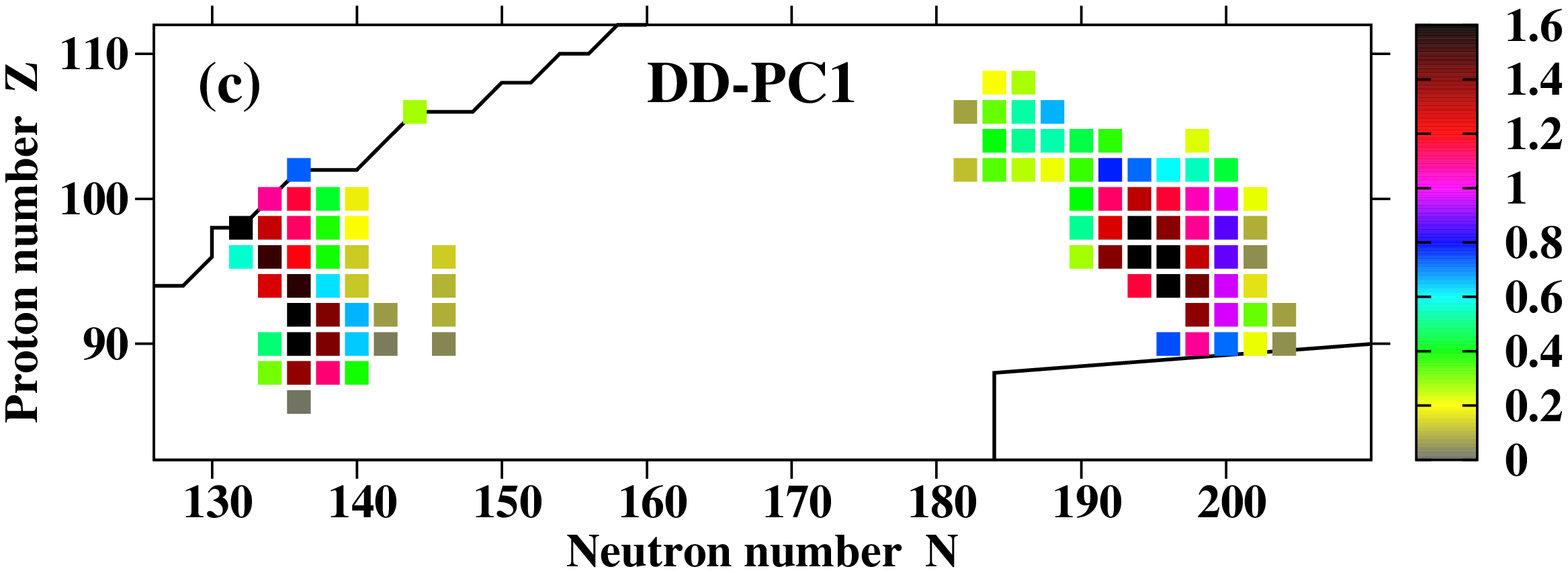}
\includegraphics[angle=-90,width=8.8cm]{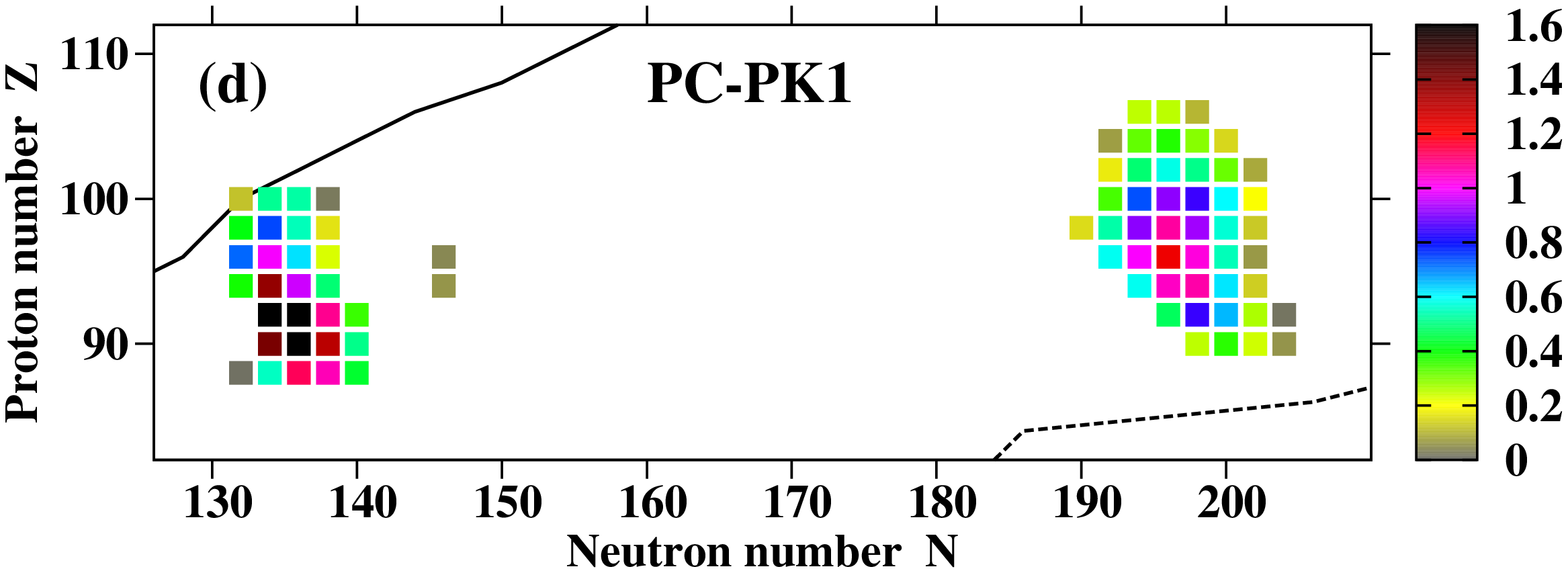}
  \caption{(Color online) Octupole deformed nuclei in the 
           selected part of nuclear chart. Only nuclei with 
           non-vanishing $\Delta E^{oct}$ are shown by squares; the 
           colors of the squares represent the values of 
           $|\Delta E^{oct}|$ (see colormap). The two-proton and 
           two-neutron drip lines are displayed by solid black 
           lines; for the CEDFs NL3*, DD-ME2 and DD-PC1 they are 
           taken from Ref.\ \cite{AARR.14}. Two-proton drip line
           for PC-PK1 is taken from Ref.\ \cite{AANR.15}. Two-neutron 
           drip line of the NL3* functional is used in panel (d) since 
           it is not defined for CEDF PC-PK1 at present.}
\label{fig-global-CDFT}
\end{figure*}

  In order to avoid the uncertainties connected with the definition of
the size of the pairing window\ \cite{KALR.10}, we use the separable form 
of the finite range Gogny pairing interaction introduced by Tian et al 
\cite{TMR.09}. Its matrix elements in $r$-space have the form
\begin{eqnarray}
\label{Eq:TMR}
V({\bm r}_1,{\bm r}_2,{\bm r}_1',{\bm r}_2') &=& \nonumber \\
= - G \delta({\bm R}-&\bm{R'}&)P(r) P(r') \frac{1}{2}(1-P^{\sigma})
\label{TMR}
\end{eqnarray}
with ${\bm R}=({\bm r}_1+{\bm r}_2)/2$ and ${\bm r}={\bm r}_1-{\bm r}_2$
being the center of mass and relative coordinates. The form factor
$P(r)$ is of Gaussian shape
\begin{eqnarray}
P(r)=\frac{1}{(4 \pi a^2)^{3/2}}e^{-r^2/4a^2}
\end{eqnarray}
The two parameters $G=728$ MeV$\cdot$fm$^3$ and $a=0.644$ 
fm of this interaction are the same for protons and neutrons and 
have been derived in Ref.\ \cite{TMR.09} by a mapping of the 
$^1$S$_0$ pairing gap of infinite nuclear matter to that of the 
Gogny force D1S~\cite{D1S}.  This pairing provides a reasonable 
description of pairing properties in the actinides (see Refs.\ 
\cite{AO.13,AARR.14,DABRS.15}) and has been used in our previous 
studies of octupole deformation in Ref.\ \cite{AAR.16}\footnote{By 
mistake the parameters $G=738$ MeV$\cdot$fm$^3$ and $a=0.636$ fm, 
derived from the D1 Gogny force \cite{TMR.09}, are quoted in 
Ref.\ \cite{AAR.16}. In reality, the same parameters $G=728$ 
MeV$\cdot$fm$^3$ and $a=0.644$ fm as the ones employed in the 
present manuscript are used in the calculations of Ref.\ 
\cite{AAR.16}.}.

  The potential energy surfaces are calculated in constrained calculations in 
the ($\beta_2,\beta_3$) plane for the $\beta_2$ values ranging from $-0.2$ up 
to 0.4 (ranging from $-0.6$ up to 0.2) if the ground state has prolate (oblate)
deformation in the calculations of Ref.\ \cite{AANR.15}) and for the $\beta_3$ 
values ranging from 0.0 up to 0.3 with a deformation step of 0.02 in each 
direction. The energies of the local minima are defined in unconstrained 
calculations.

The effect of octupole deformation can be quantitatively
characterized by the quantity $\Delta E_{oct}$ defined as
\begin{eqnarray}
\Delta E_{oct} = E^{oct}(\beta_2, \beta_3) - E^{quad}(\beta'_2,\beta'_3=0)
\end{eqnarray}
where $E^{oct}(\beta_2, \beta_3)$ and $E^{quad}(\beta'_2, \beta'_3=0)$
are the binding energies of the nucleus in two local minima of
potential energy surface; the first minimum corresponds to octupole
deformed shapes and second one to the shapes with no octupole
deformation. The quantity  $|\Delta E_{oct}|$ represents the gain of
binding due to octupole deformation. It is also an indicator of
the stability of the octupole deformed shapes. Large $|\Delta E_{oct}|$
values are typical for well pronounced octupole minima in the
PES; for such systems the stabilization of static octupole deformation
is likely. On the contrary, small $|\Delta E_{oct}|$ values are characteristic
for soft (in octupole direction) PES typical for octupole vibrations. In 
such systems  beyond mean field effects can play an important role
(see Ref.\ \cite{AAR.16}) and references quoted therein).

  The truncation of the basis is performed in such a way that all states
belonging to the major shells up to $N_F=16$ ($N_F=18$ for superheavy
$Z > 106$ nuclei) fermionic shells for the Dirac spinors and up to 
$N_B=20$ bosonic shells for the meson fields  in the case of
meson exchange functionals are taken into account. The dependence
of the calculated quantities on $N_F$ is illustrated in Fig.\ 
\ref{truncation}. One can see that all physical quantities of interest 
saturate with increasing of $N_F$. The comparison of the results shows 
that the calculations with $N_F=16$ reproduce the results of the $N_F=20$ 
truncation scheme with an accuracy of 0.007\% or better for binding 
energies, 1.6\%  for the $|\Delta E^{oct}|$ quantity, 1.56\% or better
for quadrupole deformations and 2.3\% for octupole deformation. 
Somewhat increased errors for deformations are the consequences of the 
softness of potential energy surface; for such PES some drift in the 
calculated equilibrium deformation is possible with little impact on 
total binding energy.  Note that larger basis with $N_F = 18$ is used 
for superheavy nuclei with $Z > 106$. This increase of the basis fully 
compensates the increase of the proton number in the system. As a 
result, similar or better accuracy of the description of physical 
observables is obtained in superheavy nuclei. Thus, one conclude that 
employed truncation of the basis provides sufficient numerical accuracy 
of the calculations in the vicinity of the normal deformed minimum.

\section{The properties of octupole deformed nuclei and their
         dependence on the covariant energy density functional}
\label{oct-def-dep}

  The global search for octupole deformed nuclei has been performed
for all even-even $Z=88-126$ nuclei from two-proton drip line up to 
either neutron number $N=210$ or two-neutron drip line (whichever
comes first in neutron number) employing CEDFs NL3*, DD-ME2, DD-PC1 and  
PC-PK1. Note that we use here the results obtained with CEDFs NL3* and 
DD-PC1 in Ref.\ \cite{AAR.16} for the $Z=88-106$ nuclei. Contrary to the 
results obtained within the microscopic+macroscopic approach in Ref.\ 
\cite{MNMS.95} and HFB calculations with Gogny D1S force in Ref.\ 
\cite{WE.12}, our calculations do not reveal the presence of octupole 
deformation in the ground states of superheavy nuclei with $Z\geq 110$. 
This issue will be discussed later in detail in Sec.\ \ref{uncertainties}.

\begin{figure*}
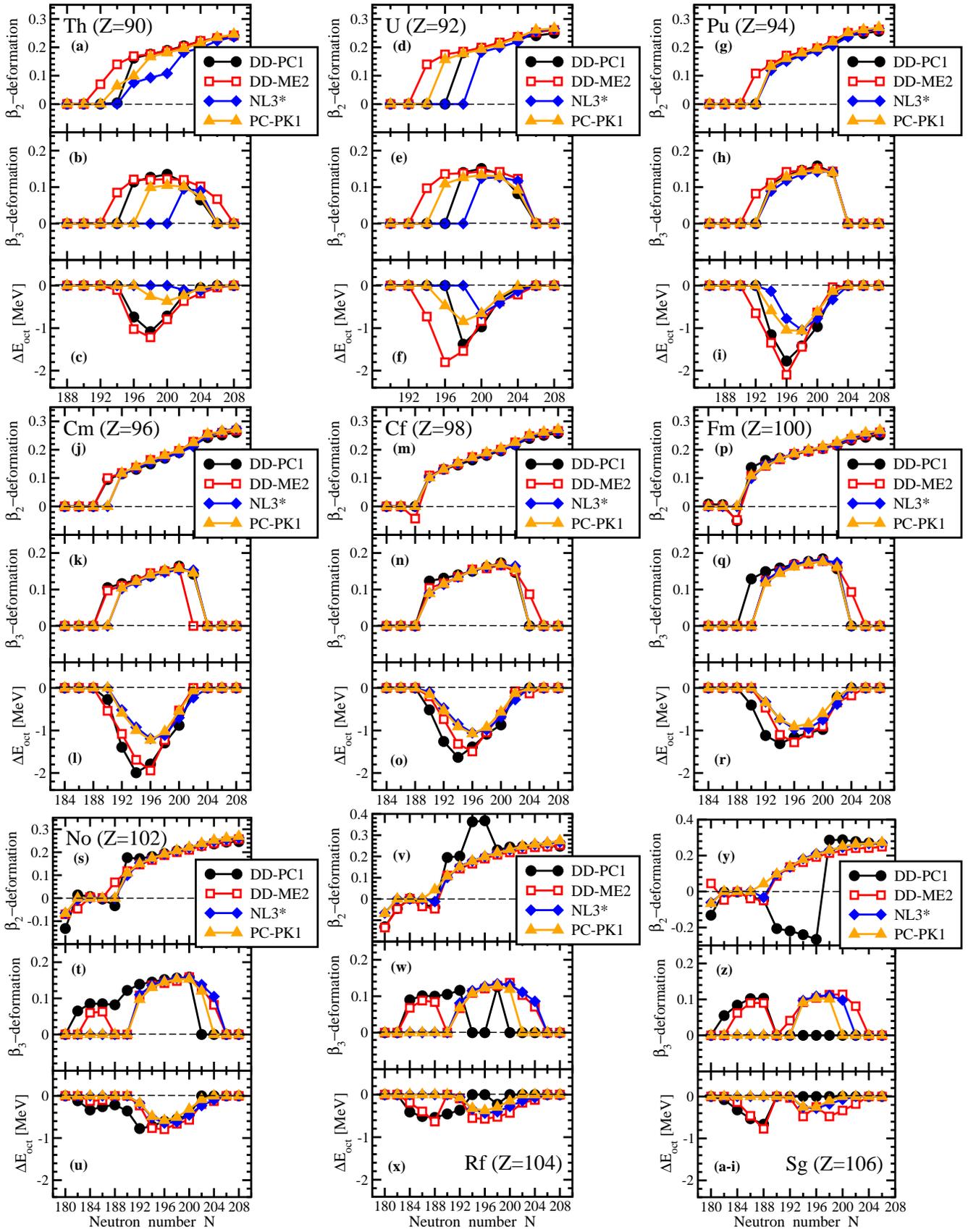

  \includegraphics[angle=0,width=5.8cm]{fig-3-a.eps}
  \includegraphics[angle=0,width=5.8cm]{fig-3-b.eps}
  \includegraphics[angle=0,width=5.8cm]{fig-3-c.eps}
  \includegraphics[angle=0,width=5.8cm]{fig-3-d.eps}
  \includegraphics[angle=0,width=5.8cm]{fig-3-e.eps}
  \includegraphics[angle=0,width=5.8cm]{fig-3-f.eps}
  \includegraphics[angle=0,width=5.8cm]{fig-3-g.eps}
  \includegraphics[angle=0,width=5.8cm]{fig-3-h.eps}
  \includegraphics[angle=0,width=5.8cm]{fig-3-m.eps}
  \caption{(Color online) The calculated equilibrium quadrupole $\beta_{2}$
  (top panel of each figure) and octupole $\beta_{3}$ (middle panel of each figure) deformations 
  as well as the $\Delta E^{oct}$ quantities (bottom panel of each figure). The employed 
  functionals are indicated.}
\label{Bet23-oct}
\end{figure*}

  Fig.\ \ref{fig-global-CDFT} shows the summary of the nuclei which possess 
octupole deformation in the ground state. The $Z \sim 92, N\sim 136$ actinides 
have been studied previously in detail in Ref.\ \cite{AAR.16} and they are shown
here only for comparison with the $Z\sim 96, N\sim 196$ region of octupole 
deformation. In both regions, the number of even-even nuclei with calculated
non-zero octupole deformation depends on the employed functional. There are 47 
(44), 57 (38), 47 (31)  and 64 (46) of such nuclei in the $Z\sim 96, N\sim 196$
($Z \sim 92, N\sim 136$) region of octupole deformation in the calculations 
with the NL3*, DD-PC1, PC-PK1 and DD-ME2 functionals, respectively. Thus, the 
calculations with CEDFs DD-ME2 and PC-PK1 confirm earlier CDFT predictions
on the existence of new region of octupole deformation centered around 
$Z\sim 96, N\sim 196$ obtained with the CEDF NL3* and DD-PC1 in Ref.\ \cite{AAR.16}.
Most of the  functionals predict that this region is substantially larger than the one 
around $Z\sim 92, N\sim 136$. Moreover, the maximum gain in binding due to octupole
deformation is comparable in the  $Z\sim 96, N\sim 196$ and  $Z\sim 92, N\sim 136$
regions. This strongly suggests the stabilization of octupole deformation in the
nuclei belonging to the central part of the $Z\sim 96, N\sim 196$ region.

  The detailed information on calculated equilibrium quadrupole ($\beta_2$) and 
octupole ($\beta_3$) deformations as well as the gains  ($\Delta E^{oct}$) in 
binding due to octupole deformation is summarized in Fig.\ \ref{Bet23-oct}. 
These results show large similarities between the NL3* 
and PC-PK1 functionals on the one hand and the DD-ME2 and DD-PC1 functionals on the
other hand. The first pair of the functionals typically shows somewhat smaller 
gain in binding due to octupole deformation as compared with second one. This is 
likely due to the fact that the pairing is stronger in neutron rich nuclei for 
the first pair of the functionals as compared with second one (see Ref.\ \cite{AA.16}); 
strong pairing leads to the reduction of $|\Delta E^{oct}|$ (see Sec. V of Ref.\ 
\cite{AAR.16}). The differences/similarities in underlying shell structure could 
be another source of observed features.

  For all functionals the maximum of the gain in binding energy 
due to octupole deformation takes place around $Z\sim 96, N\sim 196$. For nuclei in 
the vicinity of these particle numbers there is very little dependence of calculated 
equilibrium deformations on employed functional. However, on going away from these 
particle numbers the differences in calculated deformations increase because the 
nuclei become more soft in octupole deformation and thus more transitional in nature 
(see discussion in Sec.\ \ref{PES-evolution}). In particular, 
the particle numbers at which the transition from quadrupole deformed to octupole 
deformed shapes takes place become strongly dependent on the employed functional.

   Two $Z=108$ (two $Z=108$ and one $Z=110$) nuclei have non-zero octupole
deformation in the calculations with CEDF DD-PC1 (DD-ME2) 
(see Figs.\ \ref{fig-global-CDFT}b and c). They are not shown in Fig.\ 
\ref{Bet23-oct} since all these nuclei are extremely soft in octupole
deformation with very small gain in binding energy due to octupole
deformation ($|\Delta E^{oct}| <0.1$ MeV).

\section{Evolution of potential energy surfaces with particle numbers:
an example of the DD-PC1 functional.}
\label{PES-evolution}

\begin{figure*}
  \includegraphics[angle=0,width=5.9cm]{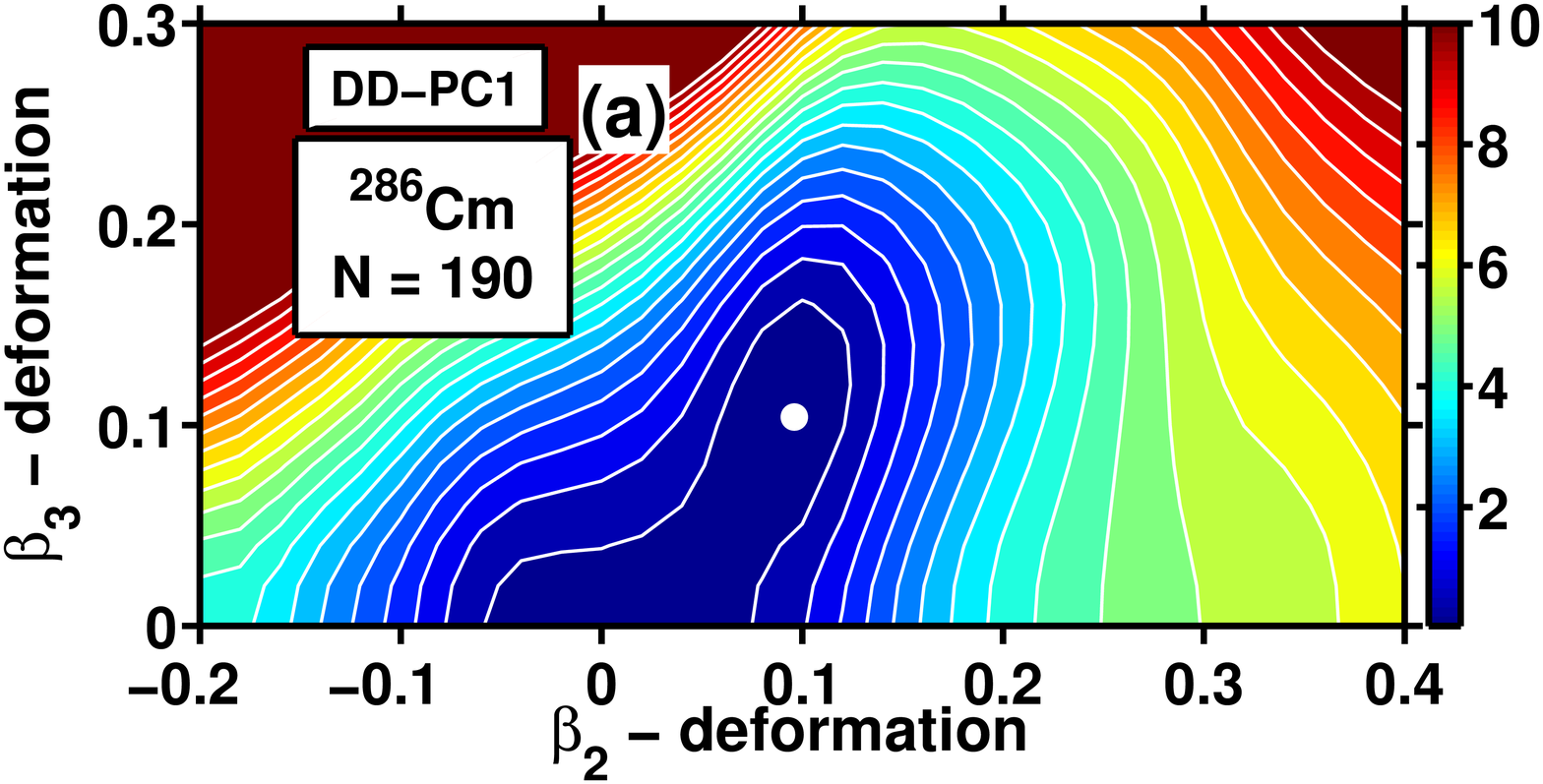}
  \includegraphics[angle=0,width=5.9cm]{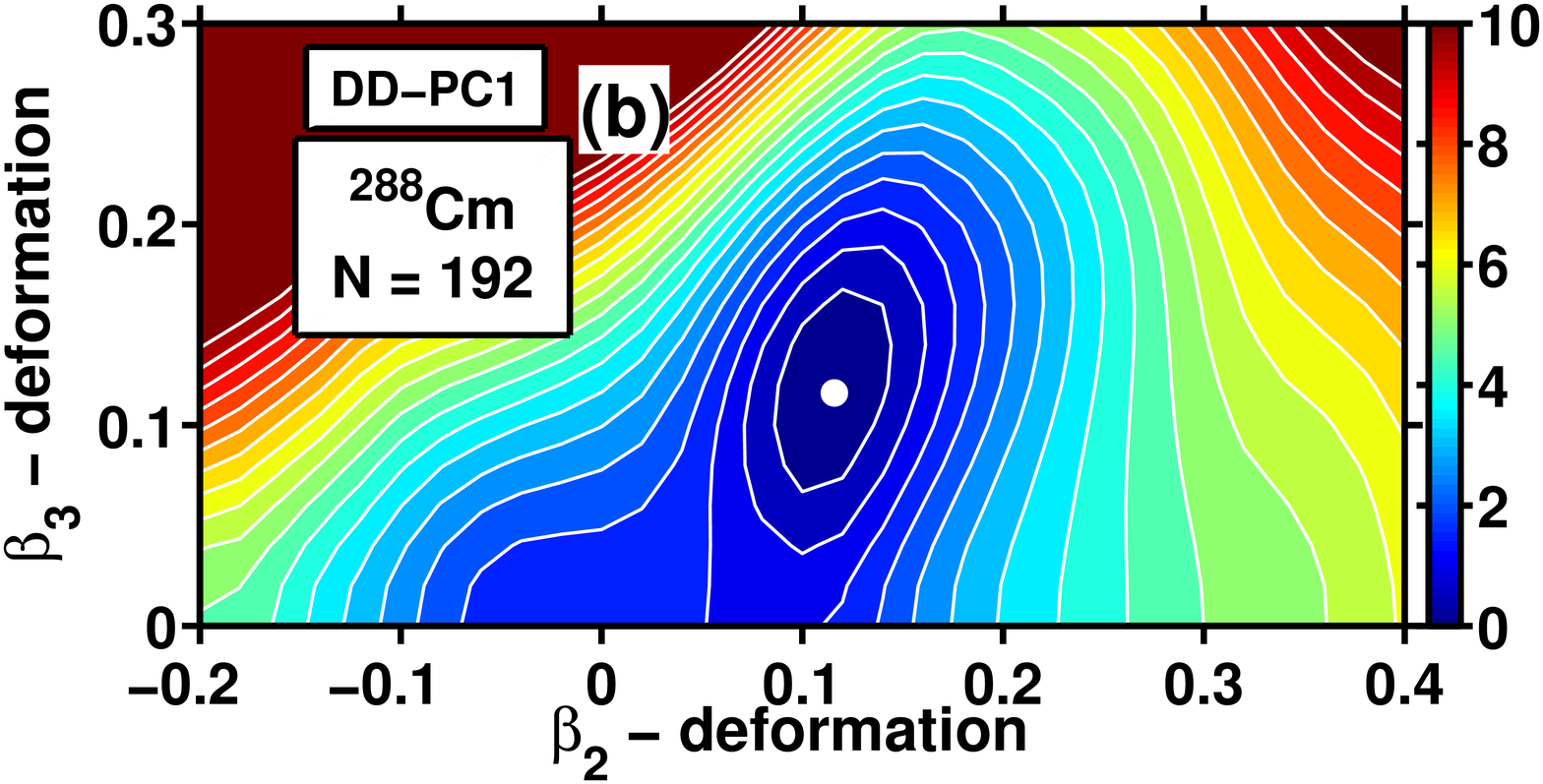}
  \includegraphics[angle=0,width=5.9cm]{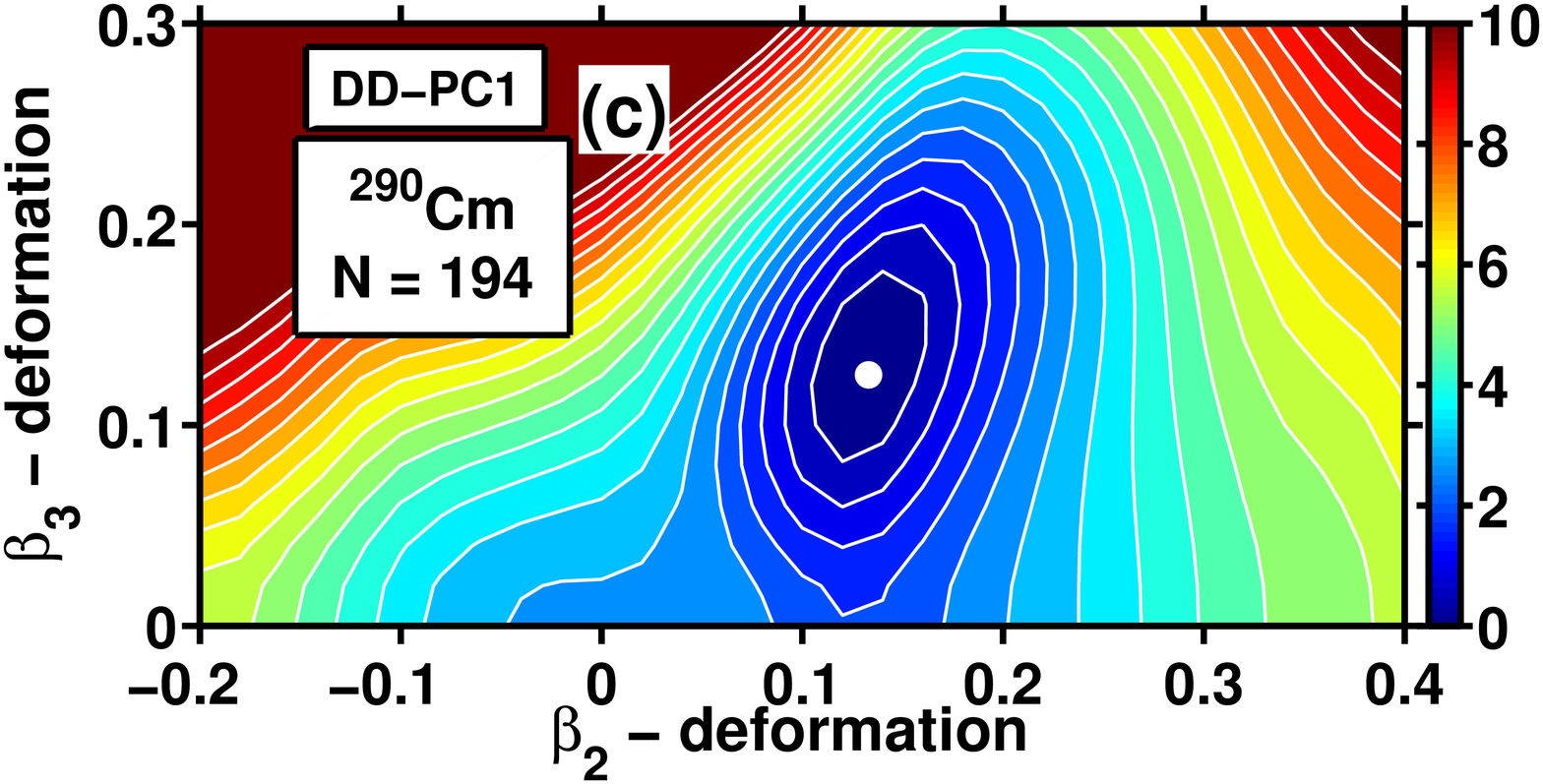}
  \includegraphics[angle=0,width=5.9cm]{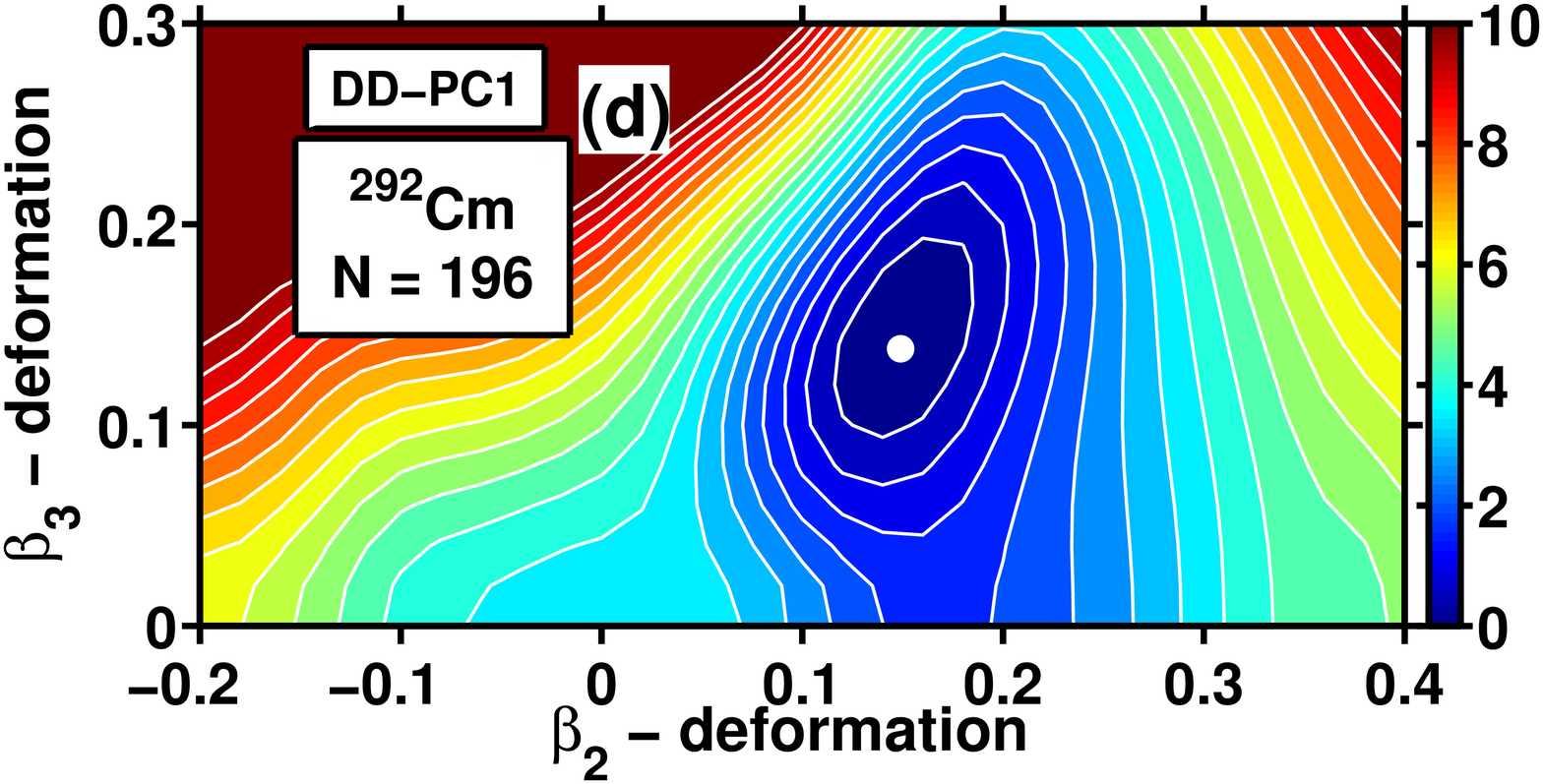}
  \includegraphics[angle=0,width=5.9cm]{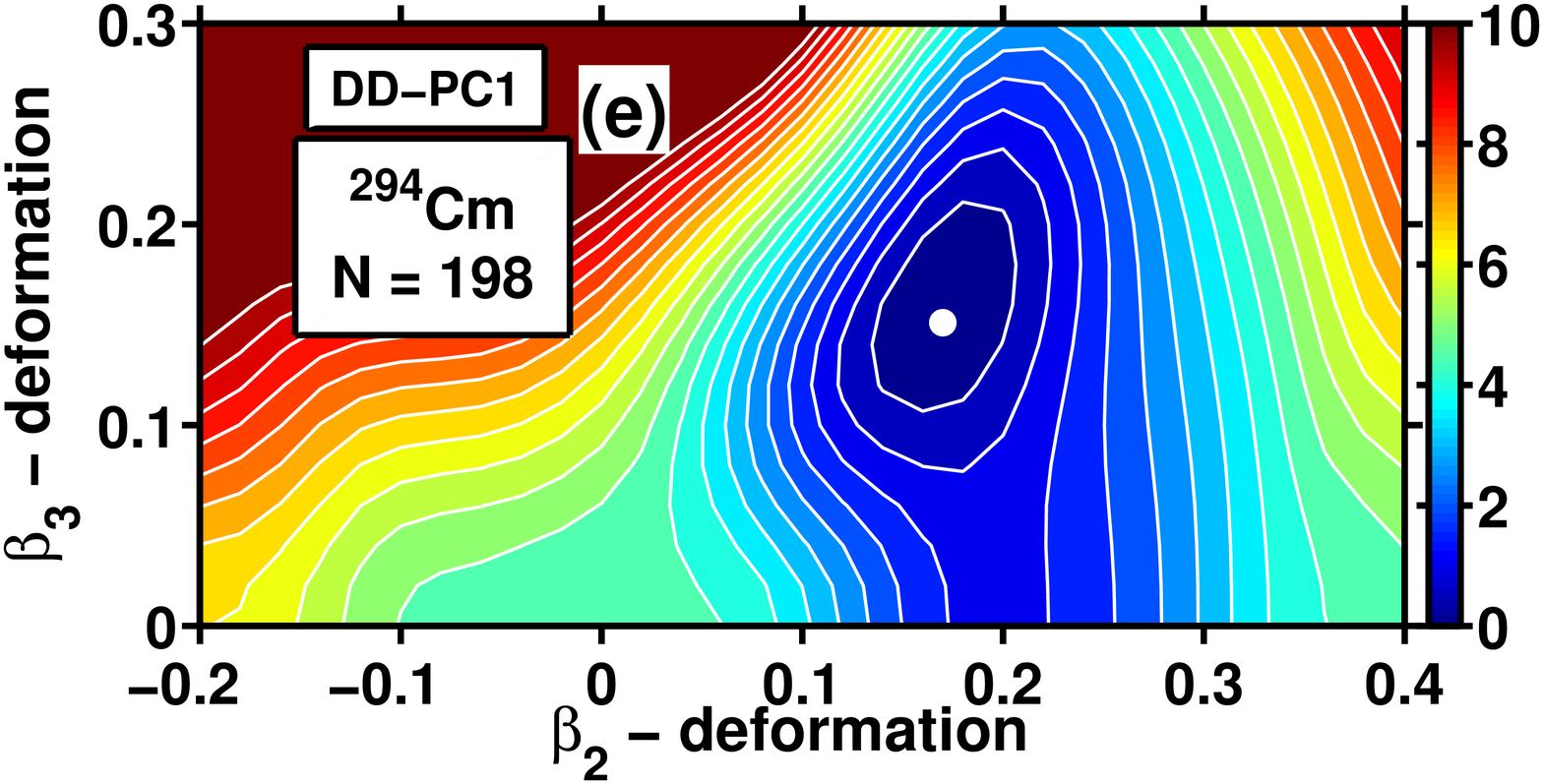}
  \includegraphics[angle=0,width=5.9cm]{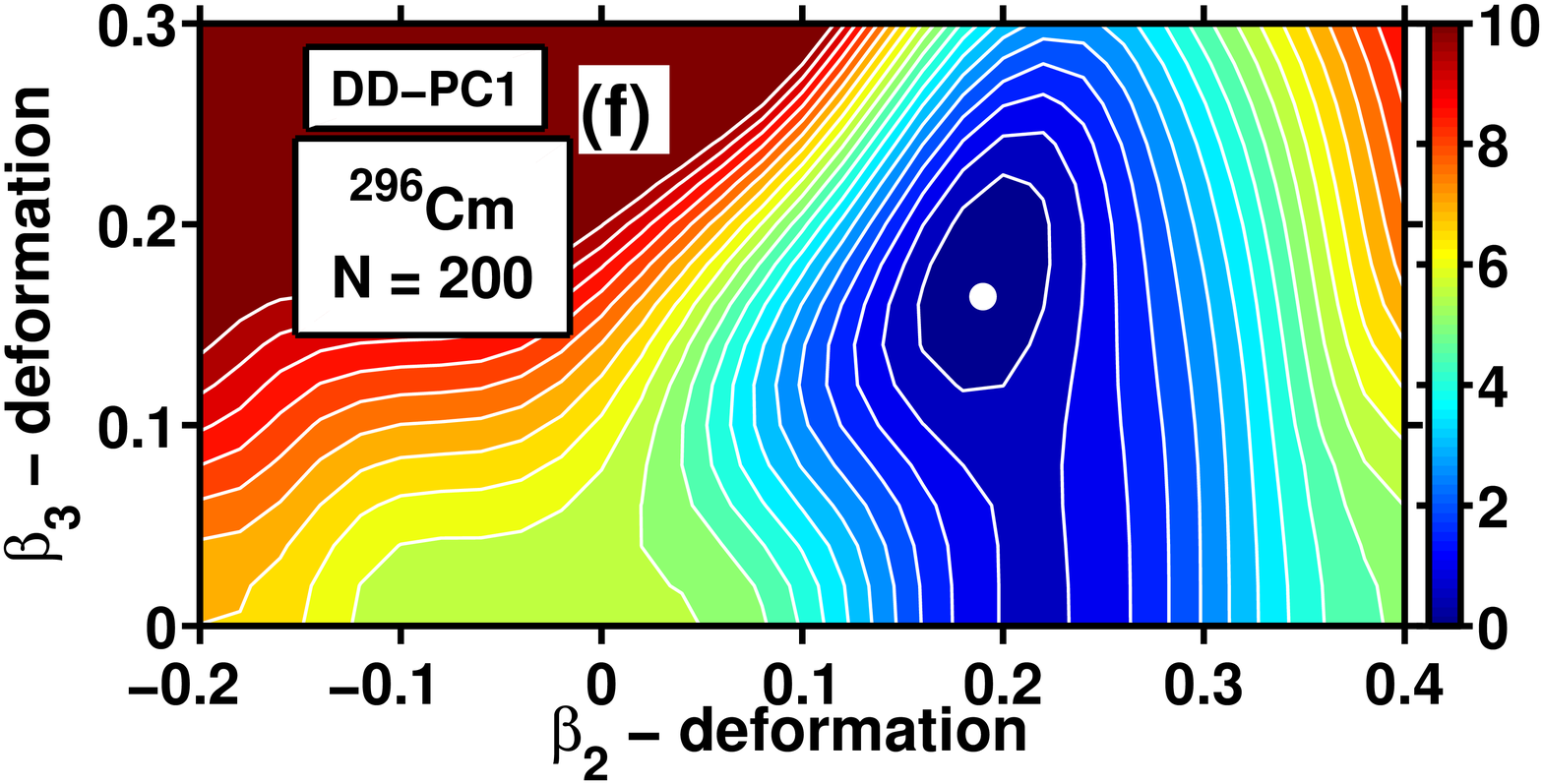}
  \includegraphics[angle=0,width=5.9cm]{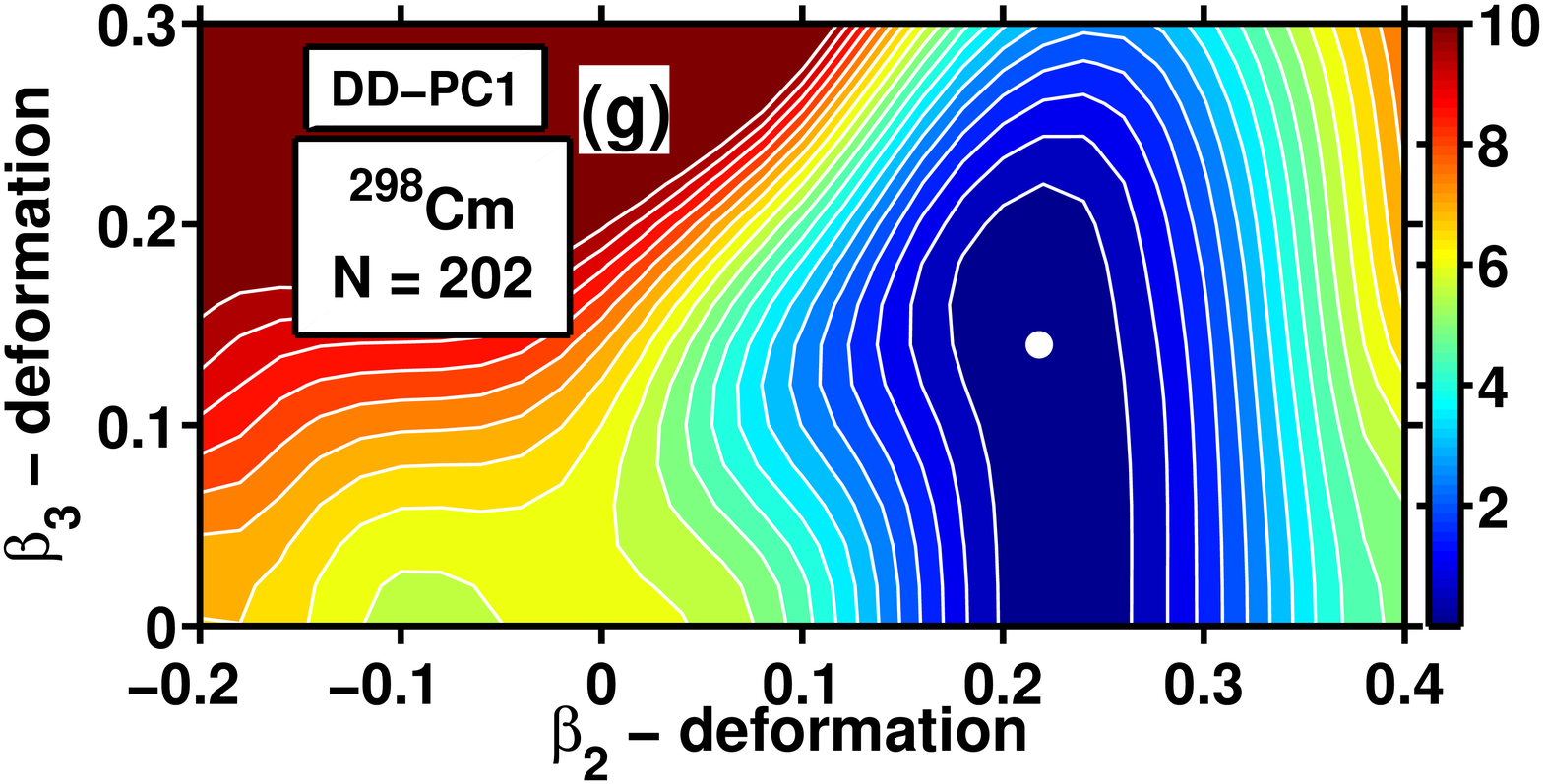}
  \caption{(Color online) Potential energy surfaces of the Cm ($Z=96$) 
          isotopes in the $(\beta_2,\beta_3)$ plane  calculated with 
          the CEDF DD-PC1. The white circle indicates the global minimum. 
          Equipotential lines are shown in steps of 0.5
          MeV. The neutron number $N$ is shown in each panel in
          order to make the comparison between different isotones
          easier.}
\label{Cm_DD-PC1}
\end{figure*}

\begin{figure*}
  \includegraphics[angle=0,width=5.9cm]{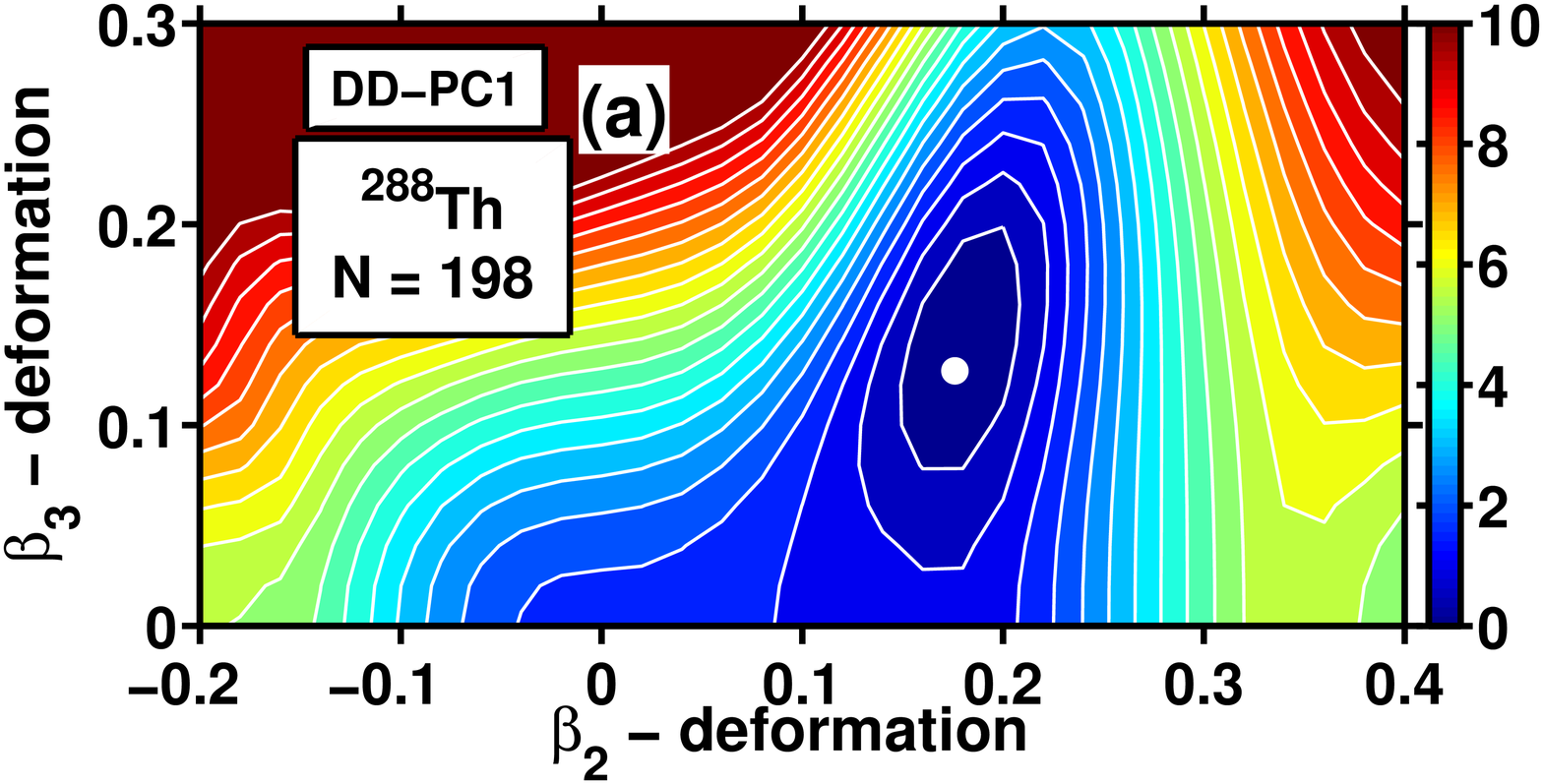}
  \includegraphics[angle=0,width=5.9cm]{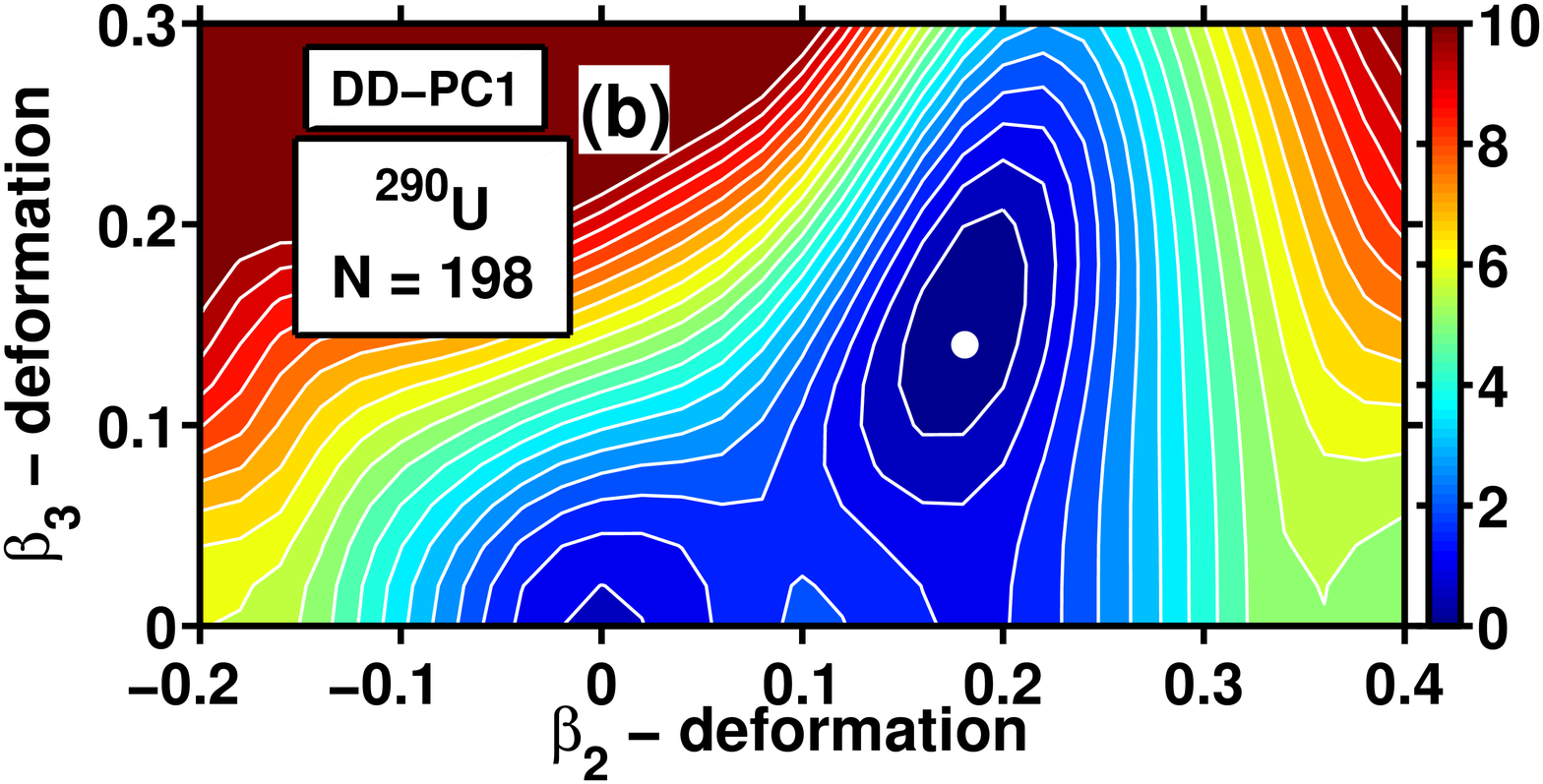}
  \includegraphics[angle=0,width=5.9cm]{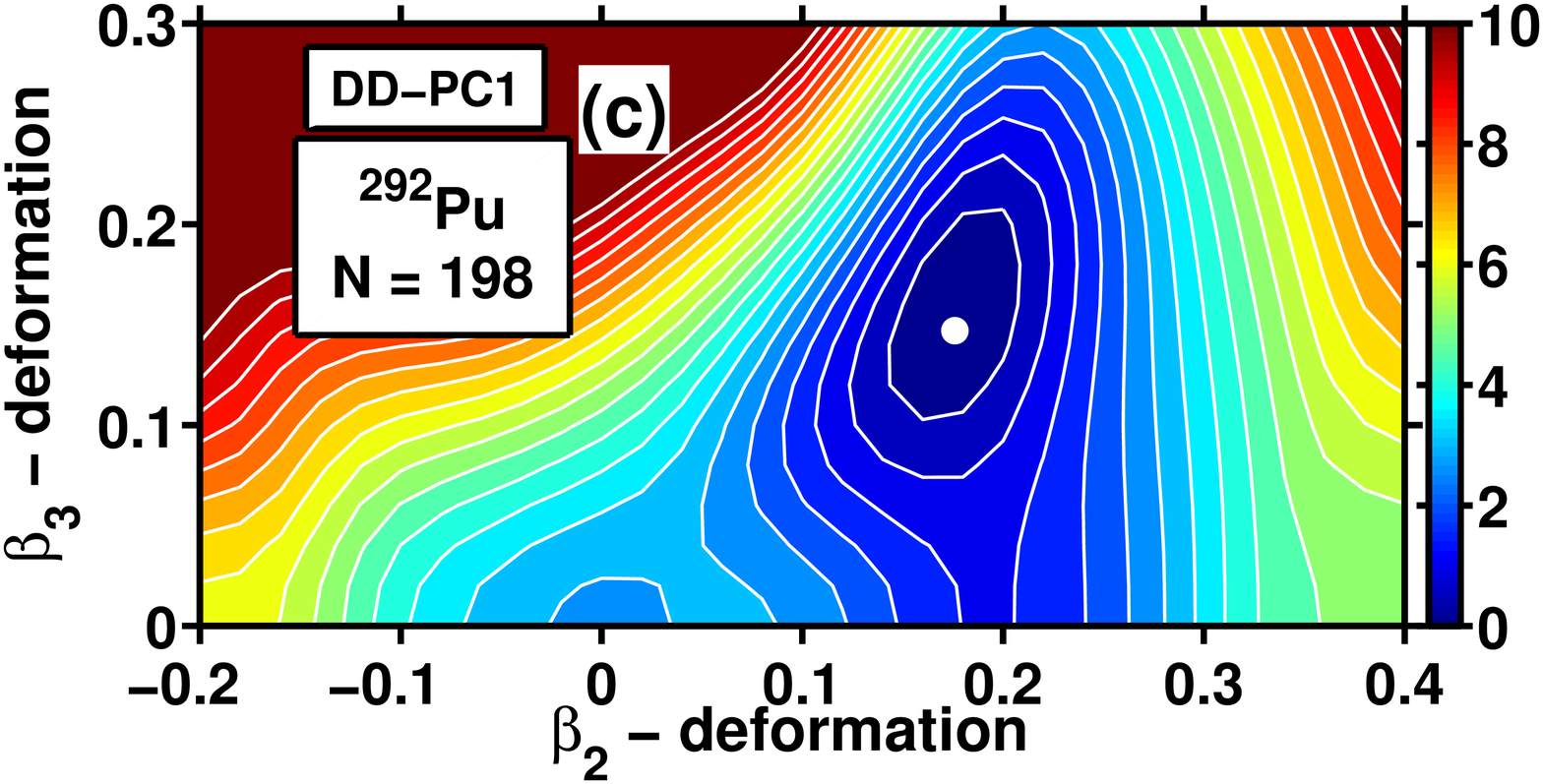}
  \includegraphics[angle=0,width=5.9cm]{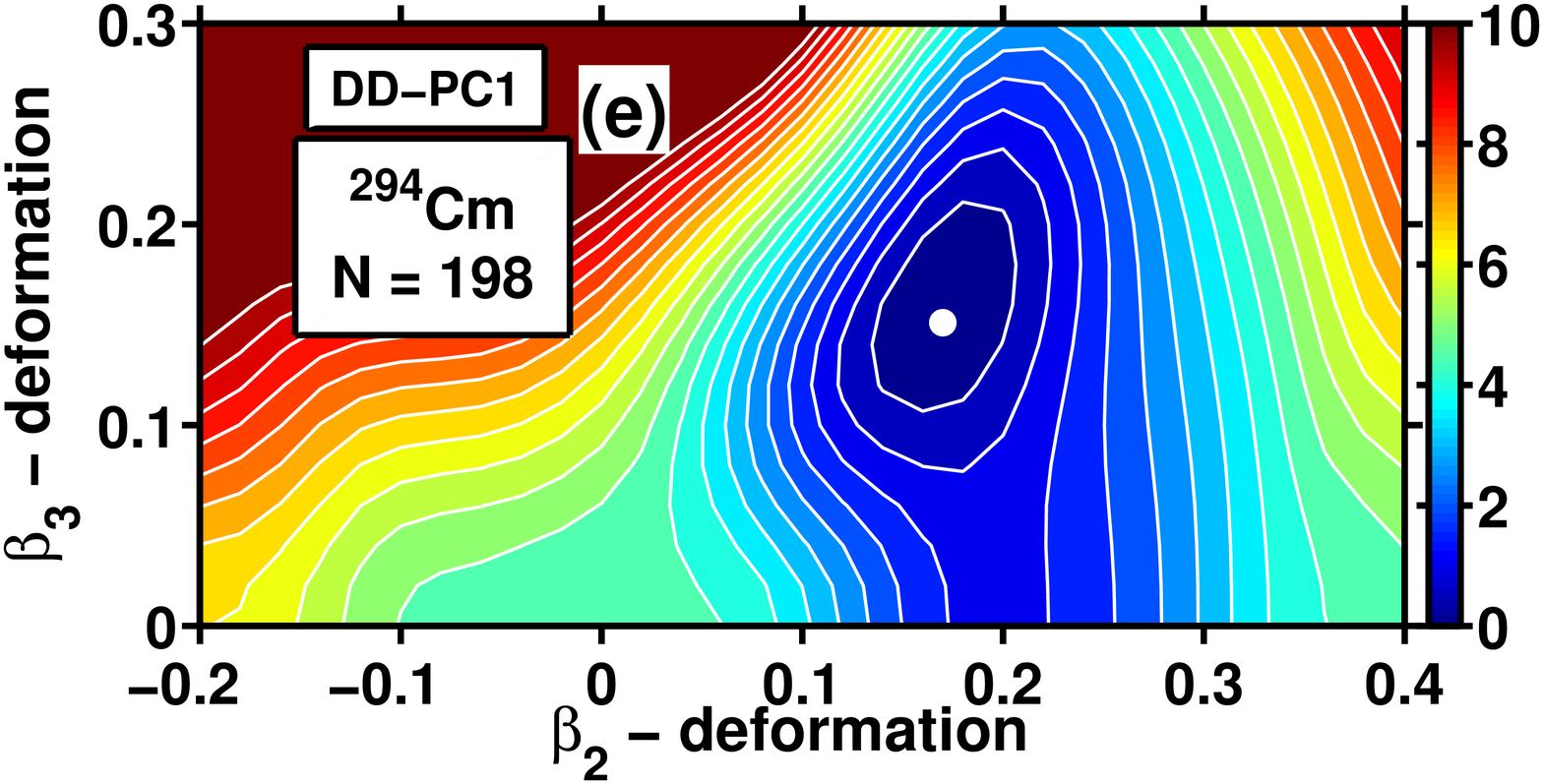}
  \includegraphics[angle=0,width=5.9cm]{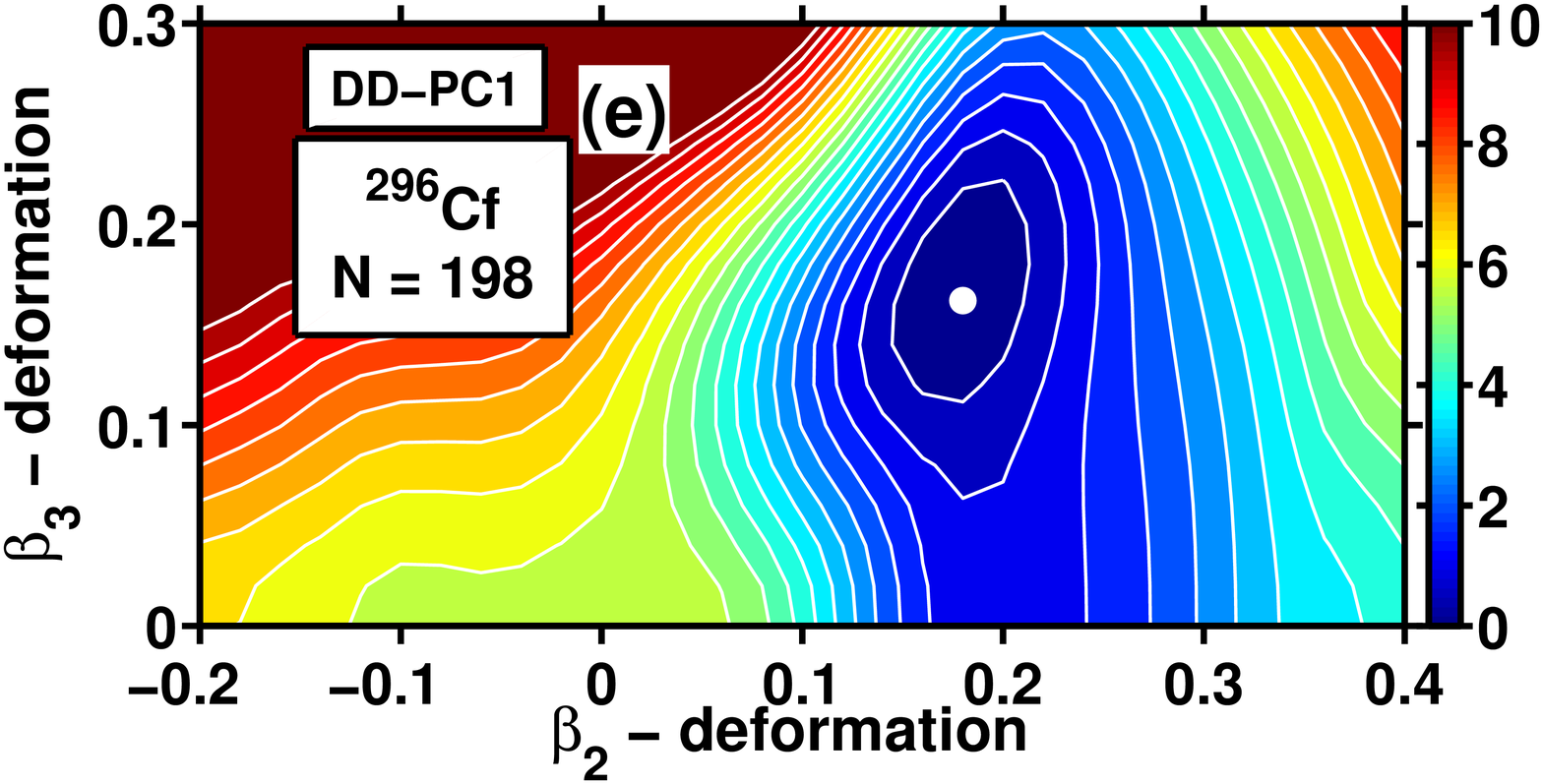}
  \includegraphics[angle=0,width=5.9cm]{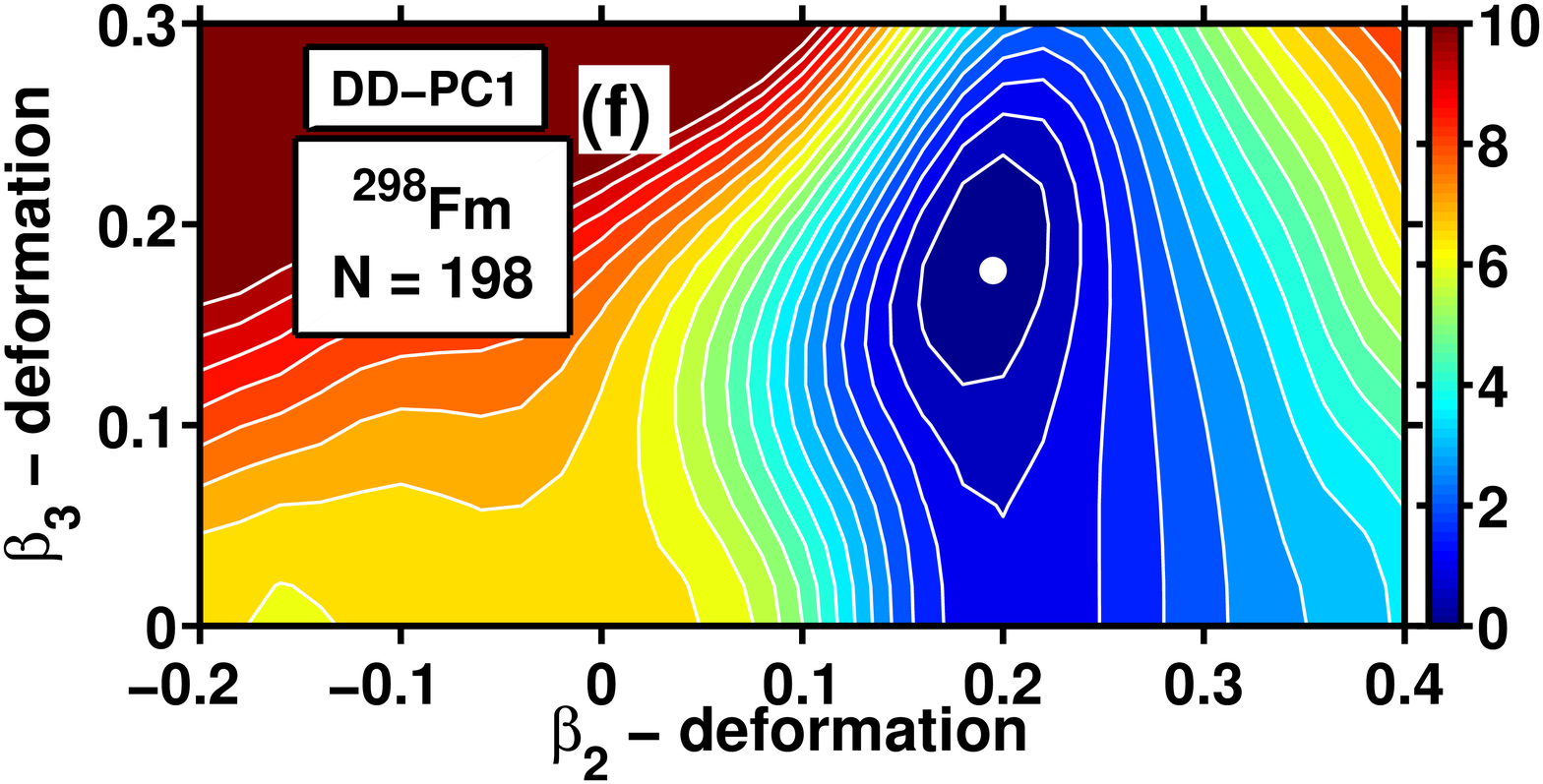}
  \includegraphics[angle=0,width=5.9cm]{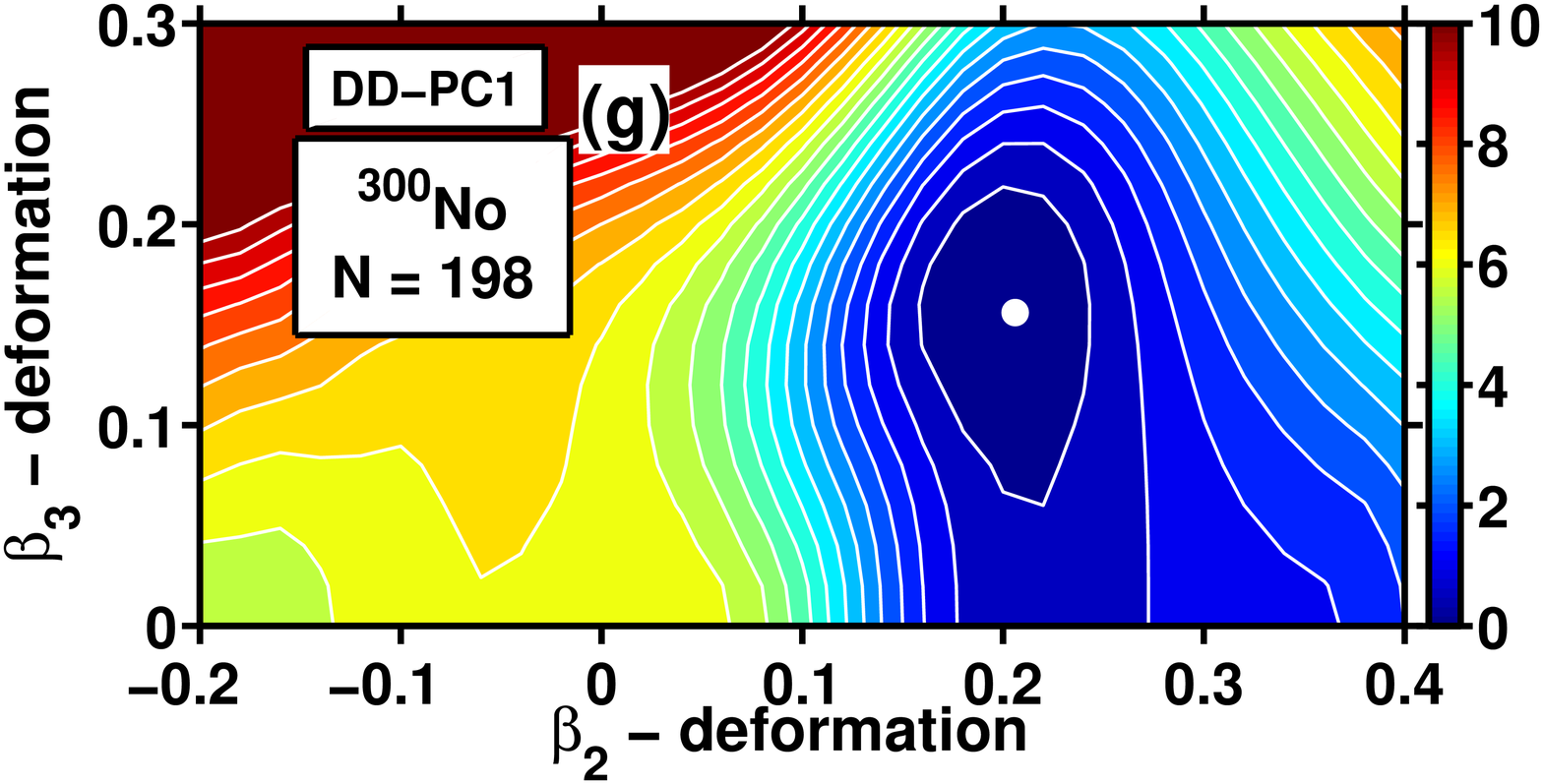}
  \includegraphics[angle=0,width=5.9cm]{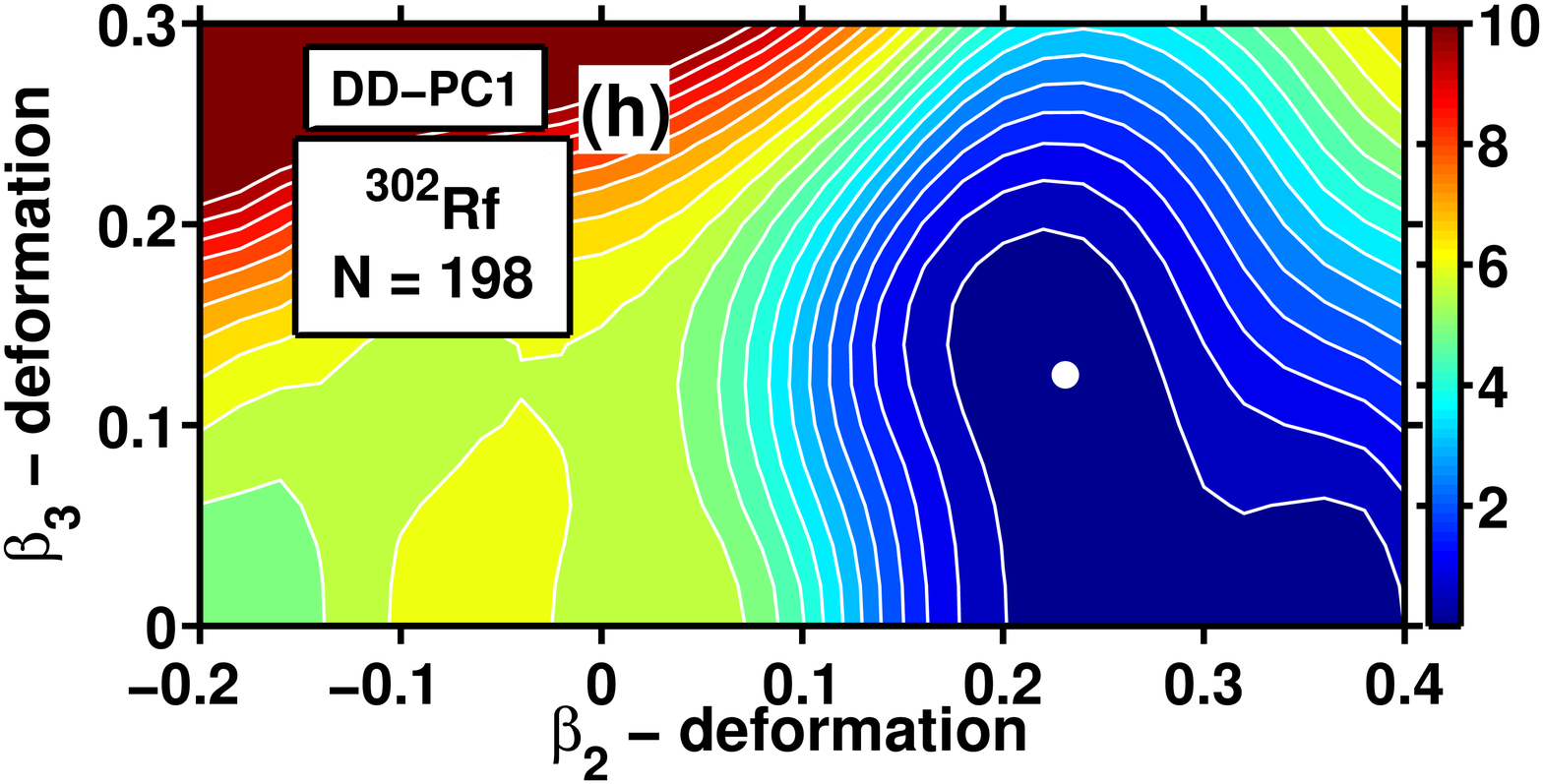}
  \caption{(Color online) The same as Fig.\ \ref{Cm_DD-PC1},
           but for the $N=198$ isotones.}
\label{N198_DD-PC1}
\end{figure*}

  In order to better understand the evolution and development of 
octupole deformation with particle number the potential energy 
surfaces (PES) of the Cm ($Z=96$) isotopes and $N=198$ isotones 
obtained in the RHB calculations with CEDF DD-PC1 are shown in Figs.\ 
\ref{Cm_DD-PC1} and \ref{N198_DD-PC1}. The center of this cross 
in the $(Z,N)$ plane represented by the $^{294}$Cm nucleus is located 
in the region of maximum gain of binding due to octupole deformation 
(see Fig.\ \ref{fig-global-CDFT}). 

  The PES of the $^{286}$Cm nucleus are rather soft in the $\beta_3$ 
direction with the gain in binding  due to octupole deformation being 
$|\Delta E_{oct}|=0.271$ MeV. The addition of the neutrons leads to the 
stabilization of octupole deformation in the $^{288-294}$Cm isotopes with 
largest gains in binding due to octupole deformation being  1.994 and 
1.790 MeV in the $^{290}$Cm and $^{292}$Cm nuclei, respectively. Subsequent 
increase of the neutron number leads to the softening of potential energy 
surfaces so that $|\Delta E_{oct}|$ is rather small (0.049 MeV) in the 
$^{298}$Cm nucleus.

  The PES for the $N=198$ isotones are displayed in Fig.\ \ref{N198_DD-PC1}. 
One can see that the lowest $Z$ nucleus ($^{288}$Th with $Z=88$) shown in this 
figure already has well pronounced minimum in octupole deformation which is 
characterized  by $|\Delta E_{oct}|=1.084$ MeV. This is because the $^{286}$Rn 
nucleus with lower $Z$ value ($Z=86$), which is expected to be more octupole 
soft, is located beyond two-neutron drip line (see Fig.\ \ref{fig-global-CDFT}). 
The $^{290}$U, $^{292}$Pu, $^{294}$Cm, $^{296}$Cf and $^{298}$Fm nuclei have well 
pronounced octupole minima in the PES. The largest gain in binding due to octupole 
deformation $|\Delta E_{oct}|=1.419$ MeV is reached  in the $^{292}$Pu nucleus. 
Subsequent increase of proton number above $Z=100$ gradually decreases 
$|\Delta E_{oct}|$ so that PES surface becomes very soft in $^{302}$Rf.

\section{Assessing systematic uncertainties in model predictions}
\label{uncertainties}

    All theoretical approaches to nuclear many body problem are based on 
some approximations. For example, in the DFT framework, there are two major 
sources of these approximations, namely, the range of interaction and the 
form of the density dependence of the effective interaction \cite{BHP.03,BB.77}. 
In the non-relativistic case one has zero range Skyrme and finite range Gogny 
forces and different density dependencies \cite{BHP.03}. A similar situation 
exists also in the relativistic case: point coupling and meson exchange models 
have an interaction of zero and of finite range, respectively 
\cite{VALR.05,DD-ME2,NL3*,DD-PC1}. The density dependence is introduced either 
through an explicit dependence of the coupling constants 
\cite{TW.99,DD-ME2,DD-PC1} or via non-linear meson couplings 
\cite{BB.77,NL3*}. This ambiguity in the definition of the range of the 
interaction and its density dependence leads to several major classes of the 
covariant energy density functionals which were discussed in detail 
in Ref.\ \cite{AARR.14}.

  These approximations lead to theoretical uncertainties in the description of
physical observables. While in known nuclei these uncertainties could be minimized
by benchmarking the model description to experimentally known nuclei (for example,
via the fitting protocol), they grow in magnitude when we extrapolate beyond known
regions \cite{DNR.14,AARR.14}. In such a situation, the estimate of theoretical 
uncertainties is needed. This issue has been discussed in detail in Refs.\ 
\cite{RN.10,DNR.14} and in the context of global studies  within CDFT in the 
introduction of Ref.\ 
\cite{AARR.14} and in Ref.\ \cite{AAR.17}. In the CDFT framework,  systematic 
theoretical uncertainties and their sources have been studied globally for the ground 
state masses, deformations, charge radii, neutrons skins, positions of drip lines etc 
in Refs.\ \cite{AARR.13,AARR.14,AARR.15,AANR.15,AAR.16,AA.16} and for inner fission 
barriers  in superheavy nuclei in Ref.\ \cite{AAR.17}.

   In the present manuscript, we focus on the uncertainties related to the 
choice of the energy density functional. Similar to our previous studies 
(\cite{AARR.14,AANR.15,AAR.16,AA.16,AAR.17}), we define systematic 
theoretical uncertainty for a given physical observable (which we call in the 
following ``spreads'')  via the spread of theoretical  predictions as 
\cite{AARR.14}
\begin{equation}
\Delta O(Z,N) = |O_{max}(Z,N) - O_{min}(Z,N)|,
\end{equation}
where $O_{max}(Z,N)$ and $O_{min}(Z,N)$ are the largest and smallest
values of the physical observable $O(Z,N)$ obtained within the set
of CEDFs under investigation for the $(Z,N)$ nucleus. 

\begin{figure*}
\includegraphics[angle=-90,width=11.8cm]{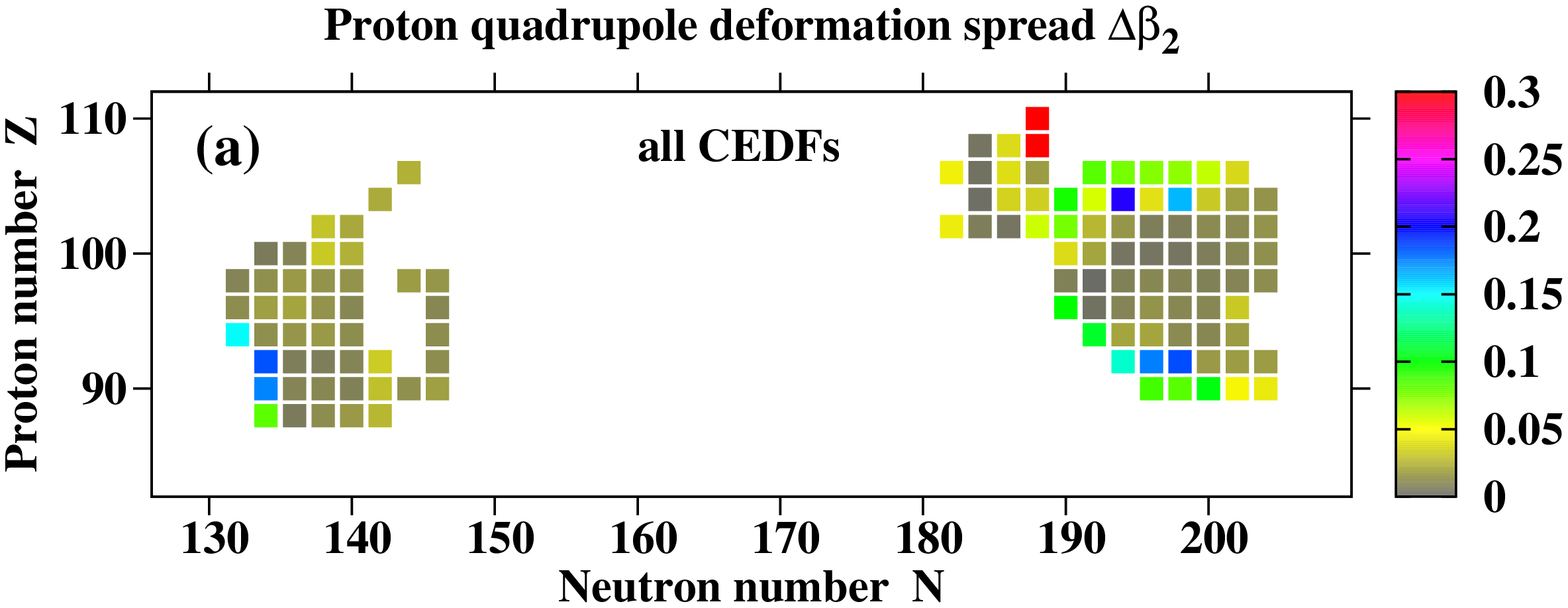}
\includegraphics[angle=-90,width=11.8cm]{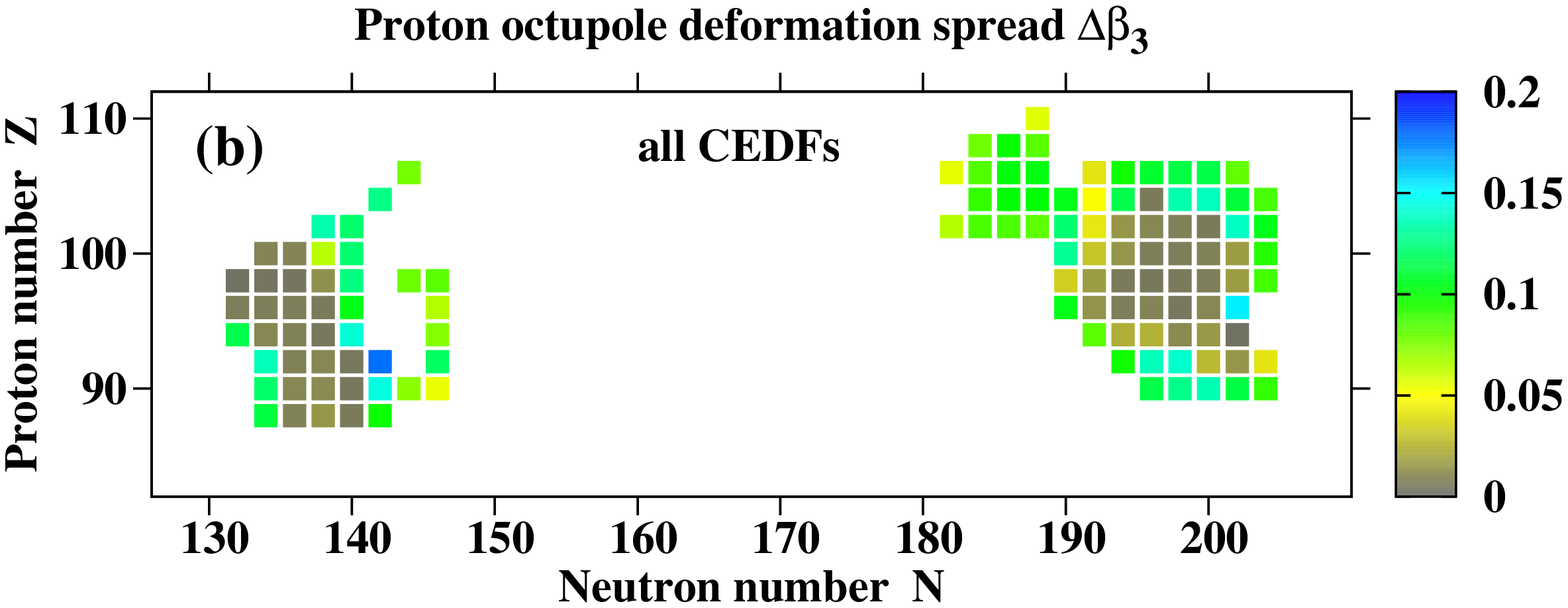}
\includegraphics[angle=-90,width=11.8cm]{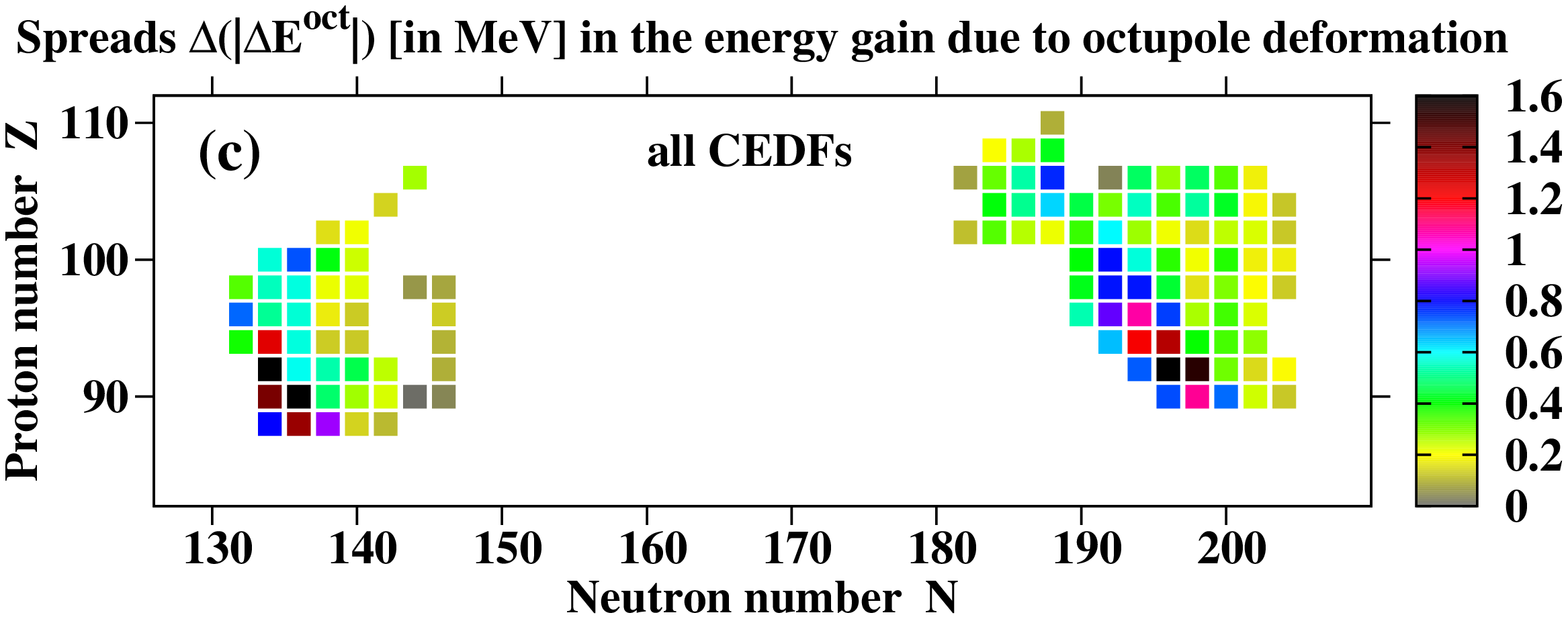}
  \caption{(Color online) The calculated spreads in quadrupole
   and octupole deformations as well as in the $|\Delta E^{oct}|$ 
   quantities. The nucleus is shown by square if it has 
   non-zero octupole deformation in the calculations with at 
   least one CEDF. }
\label{fig-spreads}
\end{figure*}

  These spreads for the calculated quadrupole and octupole deformations as 
well as for the $|\Delta E^{oct}|$ quantity are shown in Fig.\ \ref{fig-spreads}. 
One can see that the spreads for the $\beta_2$ and $\beta_3$ 
deformations in the central parts of the $Z\sim 96, N\sim 196$ and 
$Z\sim 92, N\sim 136$ regions are small. They increase at the boundaries of 
these regions where the PES of the nuclei are soft in octupole 
deformation. As a result, model predictions become strongly dependent on 
fine details of underlying single-particle structure so that the same 
$(Z,N)$ nucleus could be octupole deformed in one functional but only 
quadrupole deformed in another functional (see Fig.\ \ref{fig-global-CDFT}). Similar 
situation with low reliability of theoretical predictions in some parts of nuclear
chart has been seen earlier in the transitional regions between quadrupole deformed and 
spherical shapes (see Figs.\ 18 and 20 in Ref.\ \cite{AARR.14}) in the axial RHB 
calculations restricted to reflection symmetric shapes. The $Z=108, 110$ nuclei 
with $N=188$ show very large spreads in quadrupole
deformation (Fig.\ \ref{fig-spreads}a). These two nuclei are octupole deformed 
with $\beta_2 \sim -0.045, \beta_3 \sim 0.07$ only in the calculations with the
DD-ME2 functional. However, they are spherical in the calculations with CEDFs NL3* and
PC-PK1 but oblate (with $\beta_2 \sim -0.36$) in the calculations with DD-PC1
(see Fig.\ 6 in Ref.\ \cite{AANR.15}).

  Systematic theoretical uncertainties for the energy gain due to octupole 
deformation are shown in Fig.\ \ref{fig-spreads}c. These uncertainties show
different pattern in the $(Z,N)$ plane as compared with the uncertainties
for the $\beta_2$ and $\beta_3$ deformations (Figs.\ \ref{fig-spreads}a and b).
The maximum uncertainties for the $|\Delta E^{oct}|$ quantity exists in the
left bottom corners of the $Z\sim 96, N\sim 196$ and $Z\sim 92, N\sim 136$
regions of octupole deformation. Theoretical uncertainties gradually decrease 
on going away from these corners and become quite small at the boundaries of 
the regions of octupole deformation. This is not surprising considering the 
fact that the nuclei at these boundaries are octupole soft with rather small 
gain in binding due to octupole deformation. 

\begin{figure*}
\includegraphics[angle=-90,width=8.8cm]{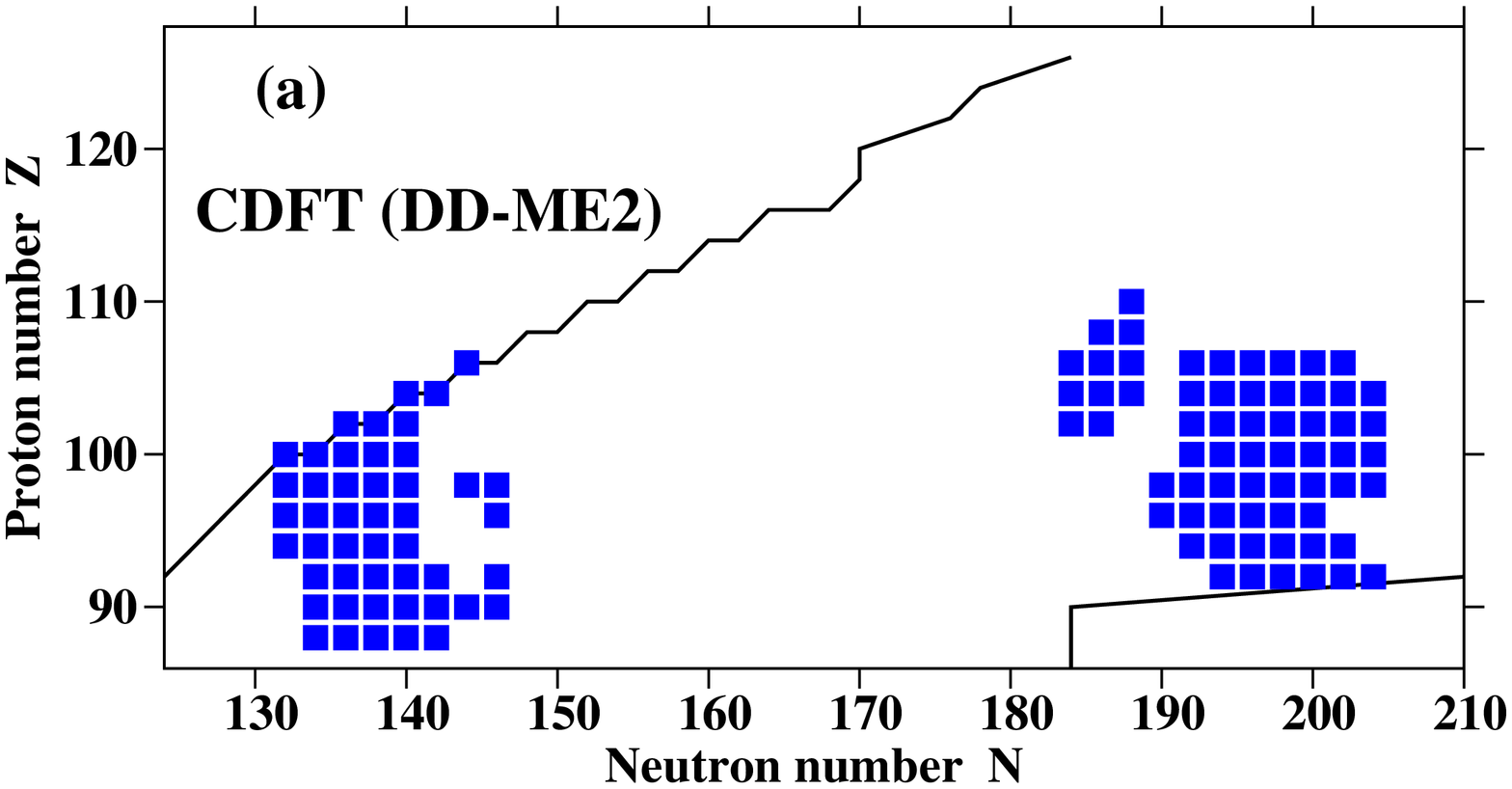}
\includegraphics[angle=-90,width=8.8cm]{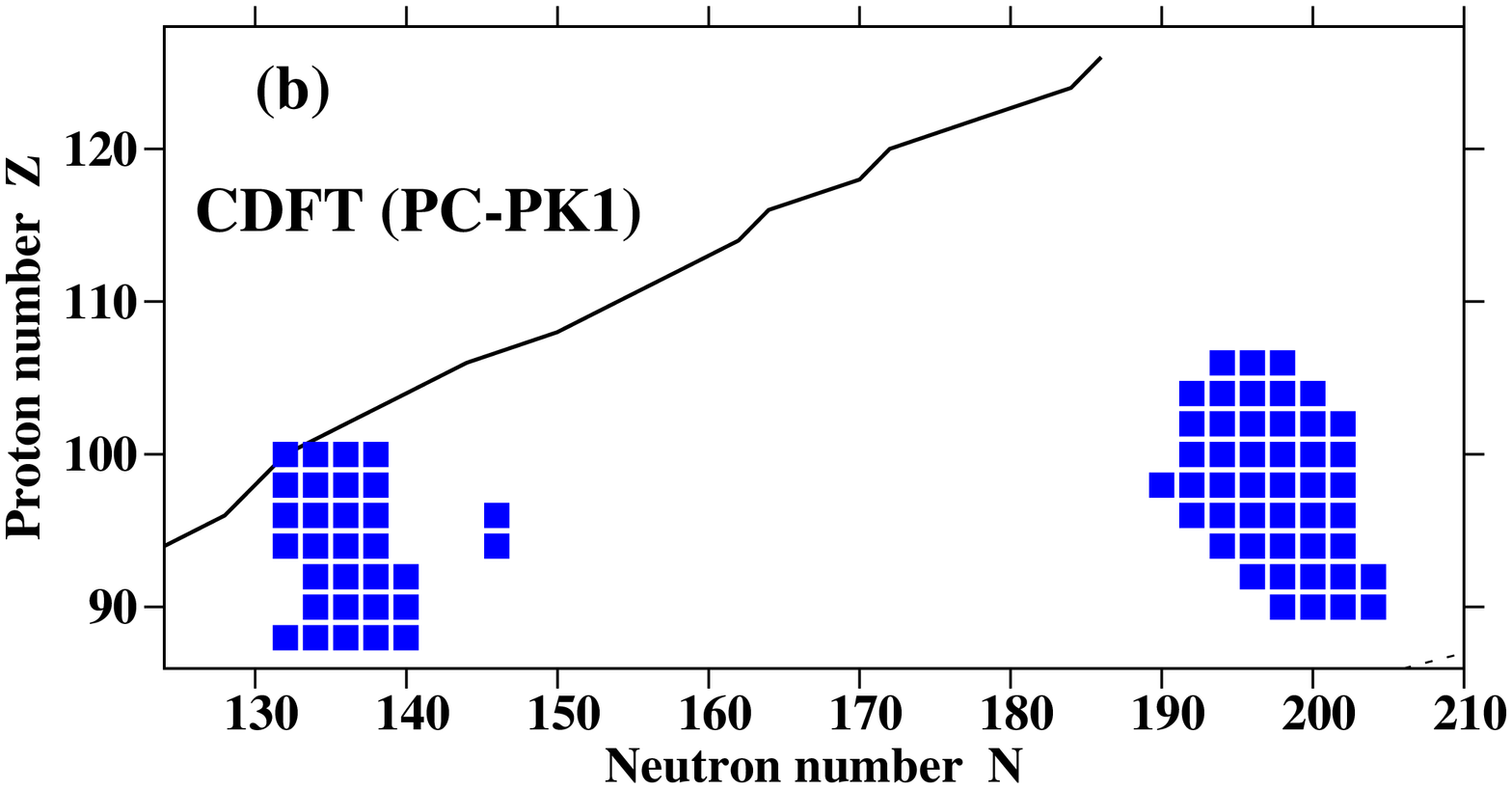}
\includegraphics[angle=-90,width=8.8cm]{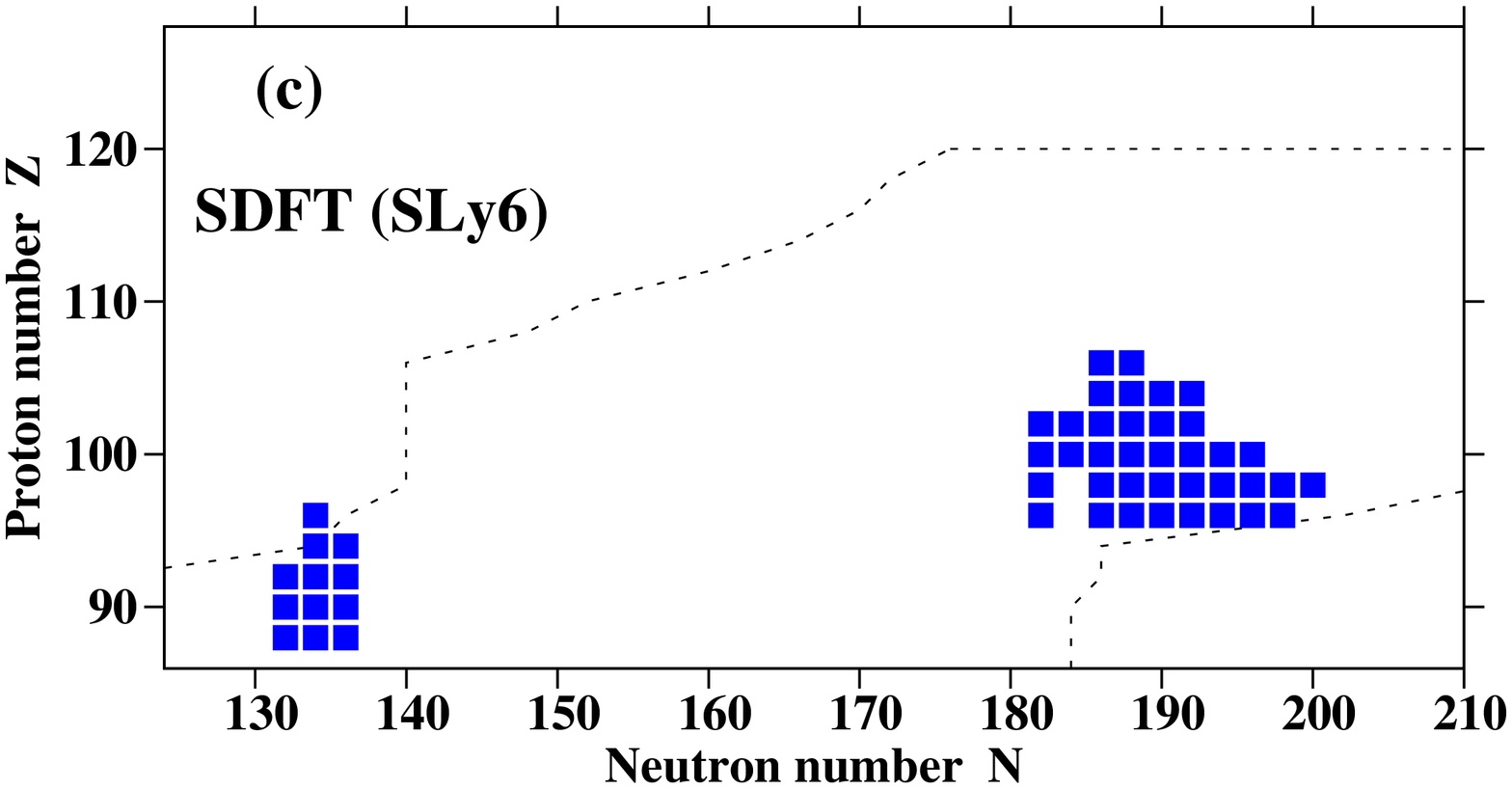}
\includegraphics[angle=-90,width=8.8cm]{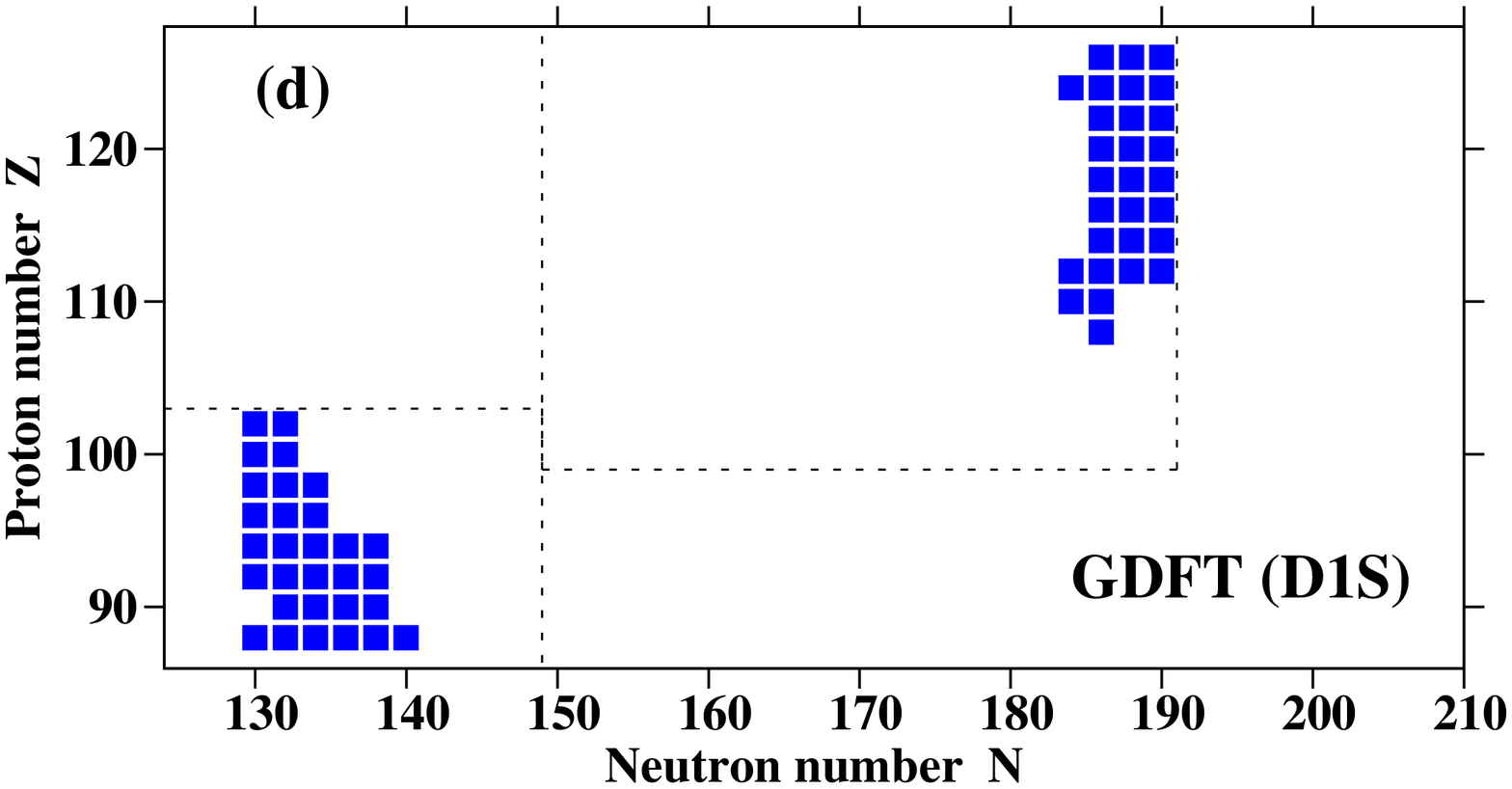}
\includegraphics[angle=-90,width=8.8cm]{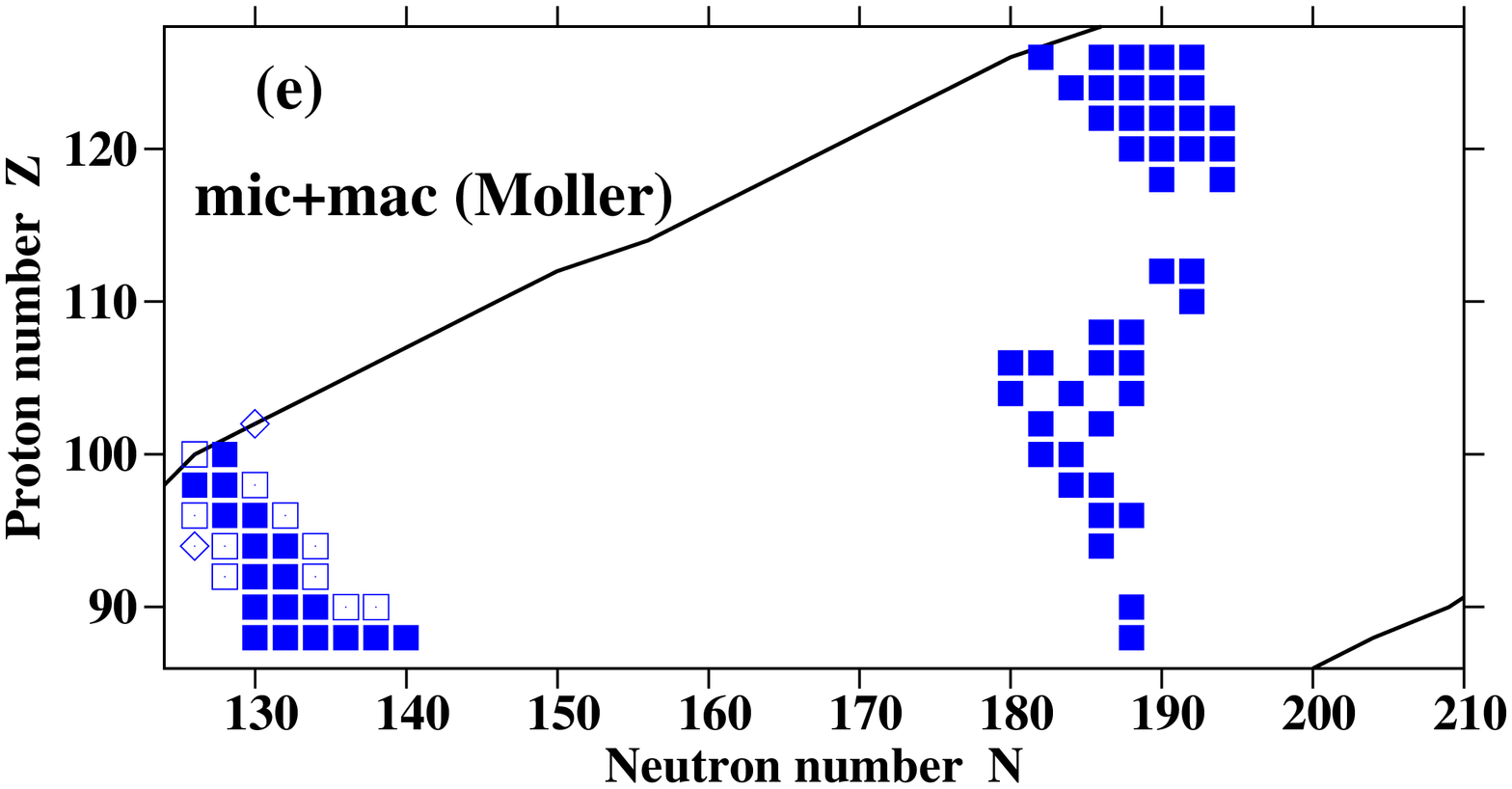}
  \caption{(Color online) Octupole deformed nuclei obtained in the CDFT 
          calculations with CEDFs DD-ME2 (panel (a)) and PC-PK1 (panel (b)), 
          in Skyrme DFT calculations with SLy6 functional \protect\cite{ELLMR.12} 
          (panel (c)), Gogny DFT calculations with EDF D1S \protect\cite{RR.12,WE.12} 
          (panel (d))  and in microscopic+macroscopic calculations of Ref.\ 
          \protect\cite{MNMS.95,MBCOISI.08} (panel (d)).  Only nuclei with non-zero 
          calculated octupole deformation are shown by squares. The two-proton 
          and two-neutron drip lines are displayed by solid black lines in panels 
          (a), (b) and (e).  In panels (c) and (d), the regions in which the 
          searches for octupole deformation have been performed are outlined by 
          dashed lines. The results of the SDFT calculations for the 
          $Z\sim 100, N\sim 190$ region of octupole
          deformation are extracted from Fig.\ 11 of Ref.\ \protect\cite{ELLMR.12}.
          The $Z\sim 92, N\sim 134$ region of octupole deformation in 
          panel (c) is shown schematically (based on Fig.\ 4 of Ref.\ 
          \protect\cite{ELLMR.12}). Note that the results presented in panel (d) in
          two regions of octupole deformation have been obtained in two independent 
          calculations of Refs.\ \protect\cite{RR.12,WE.12}.
          The nuclei which are octupole deformed in the mic+mac calculations
          of Ref.\ \cite{MNMS.95} are shown by solid blue squares and open diamonds 
          in panel (d).  Open squares indicate additional (as compared with Ref.\ 
          \cite{MNMS.95}) octupole deformed nuclei obtained in Ref.\ \protect\cite{MBCOISI.08},
          while open diamonds the nuclei which cease to be octupole deformed 
          (as compared with Ref.\ \protect\cite{MNMS.95}) in the mic+mac calculations of
          Ref.\ \protect\cite{MBCOISI.08}.
}
\label{fig-global}
\end{figure*}

  It is important to compare the CDFT predictions for 
the $Z\sim 96, N\sim 196$ region of octupole deformation with the 
ones obtained in non-relativistic theories. Such a comparison is 
presented in Fig.\ \ref{fig-global} where two extreme CDFT predictions 
for octupole deformed region [the largest (smallest) $Z\sim 96, N\sim 
196$ region of octupole deformation is obtained in the calculations 
with DD-ME2 (PC-PK1) functional] in the indicated part of nuclear chart 
are compared with the predictions obtained in the Skyrme and Gogny 
DFTs and macroscopic+microscopic approach. The Skyrme DFT calculations 
with the SLy6 functional predict such a region with the center 
located around $Z=100, N=190$ \cite{ELLMR.12} (see Fig.\ 
\ref{fig-global}c). Similar region of octupole deformation (but with 
smaller gain in binding energy due to octupole deformation) has also 
been obtained in the calculations with the SV-min EDF \cite{ELLMR.12}. 
The Gogny DFT calculations are limited in the $(Z,N)$ plane (see Fig.\ 
\ref{fig-global}d); even then they do not indicate the presence of 
octupole deformation in the nuclei located in the upper parts of
the regions of octupole deformation obtained
in the Skyrme and CDFT calculations. However, the extension of the Gogny DFT 
calculations to the $Z=90-108, N=180-210$ region of nuclear chart 
is needed to clarify the question of the existence of the $Z\sim 96, 
N\sim 196$ region of octupole deformation in this type of the EDFs. On 
the contrary, the mic+mac calculations of Ref.\ \cite{MNMS.95} predict 
the existence of octupole deformation in this region (Fig.\ 
\ref{fig-global}e). However, the island of octupole deformation
is smaller than the one obtained in the CDFT or Skyrme DFT 
calculations and it is centered around $Z=100, N=184$. It is 
necessary to mention that that these results have been obtained
more than twenty years ago. Newer mic+mac calculations of Ref.\ 
\cite{MBCOISI.08} do not cover this part of nuclear chart. However, 
in the $Z\sim 92, N\sim 134$ region of octupole deformation, the 
number of octupole deformed even-even nuclei is increased from 20 
in Ref.\ \cite{MNMS.95} to 27 in Ref.\ \cite{MBCOISI.08}. It would be 
interesting to see how the number of octupole deformed nuclei in 
the $Z\sim 100, N\sim 184$ region would be modified if newer formalism 
of mic+mac approach of Ref.\ \cite{MBCOISI.08} with improved model
parameters would be applied to this region.

  Although placing the center of the island of octupole deformed nuclei at 
different particle numbers (at $Z\sim 96, N\sim 196$ in CDFT, at $Z\sim 100, 
N\sim 190$ in Skyrme DFT and at $Z\sim 100, N\sim 184$ in mic+mac approach), 
modern theories agree on the existence of such island in neutron-rich actinides 
and low-$Z$ superheavy nuclei. However, their predictions diverge for the 
$Z \geq 110$ superheavy nuclei. The CDFT calculations of the present
manuscript and the Skyrme DFT calculations of Ref.\ \cite{ELLMR.12} do not 
predict the existence of octupole deformation in the ground states of the 
$110 \leq  Z \leq 126$ and $110 \leq  Z \leq 120$ superheavy nuclei, respectively.
On the contrary, the Gogny DFT (Fig.\ \ref{fig-global}d and Ref.\ \cite{WE.12}) 
and mic+mac (Fig.\ \ref{fig-global}e and Ref.\ \cite{MNMS.95}) calculations 
predict the existence of such nuclei. The HFB calculations 
based on the Gogny D1S force predict octupole deformation in the ground states of 
the $(Z=108-126, N=186-190)$ even-even nuclei (see Fig.\ 3 in Ref.\ \cite{WE.12}).
These nuclei either do not have quadrupole deformation (the $N=186$ and some 
$N=188$ nuclei) or this deformation is rather small ($\beta_2 <0.1$) for 
$N=190$ and some $N=188$ nuclei. The octupole deformation is rather small
for most of these nuclei apart of few $N=188$ nuclei and the majority of the 
$N=190$ nuclei which have substantial octupole deformation $\beta_3$ 
exceeding  0.1. Note that these calculations cover only nuclei with  $N\leq 190$. 
More extensive mic+mac calculations of Ref.\ \cite{MNMS.95} indicate larger 
region of octupole deformation in the superheavy nuclei (see Fig.\ \ref{fig-global}e).

 The existence of octupole deformed shapes is dictated by the underlying shell 
structure. Strong octupole coupling exists for particle numbers associated with 
a large $\Delta N=1$ interaction between intruder orbitals with $(l,j)$ and 
normal-parity orbitals with $(l-3,j-3)$ \cite{BN.96}. Thus, the discussed above 
differences in the model predictions are traced back to the differences
in the underlying single-particle structure. For normal deformed nuclei not far 
away from beta stability the tendency towards octupole deformation or strong 
octupole correlations occurs just above closed shells. For example, in the CDFT 
the maximum of octupole correlations takes place in the $A\sim 230$ region of 
octupole deformation at proton number $Z\sim 92$ (the coupling  between the proton 
1$i_{13/2}$ and  2$f_{7/2}$ orbitals) and 136 (the coupling  between the neutron 
1$j_{15/2}$ and 2$g_{9/2}$ orbitals).  In the $Z\sim 96, N\sim 196$ region, the 
presence of octupole deformation is due to the interaction of the 2$h_{11/2}$ 
and  1$k_{17/2}$ neutron orbitals and of the 1$i_{13/2}$ and 2$f_{7/2}$ proton 
orbitals. Note that the maximum of the interaction of proton orbitals occurs 
at a higher proton number $Z$ as compared with the well known $A\sim 230$ region 
of octupole deformation in actinides.  In the $Z\sim 120, N\sim 190$ region,
the interaction of the 2$h_{11/2}$ and  1$k_{17/2}$ neutron orbitals and 
of the 1$j_{15/2}$ and 2$g_{9/2}$ proton orbitals are responsible for 
strong octupole correlations in the Gogny DFT and mic+mac calculations. 
However, the energies of these states and their positions with respect
of the Fermi level are described differently in different models (see, for 
example, Figs. 1, 4, 9 and 15 in Ref.\ \cite{BRRMG.99}, Fig.\ 4 in Ref.\ 
\cite{BNR.01}, and Fig. 1 in Ref.\ \cite{AANR.15}).

  The predictive power of above discussed models in the description of these
energies and, as a consequence, of the regions of octupole deformation 
decreases on going away from known region of nuclear chart. Some differences 
in the predictions of the region of octupole deformation do already exist 
for known $A \sim 230$ region of octupole deformation (see Fig.\ \ref{fig-global} 
and discussion in Ref.\ \cite{AAR.16}). However, they become magnified with 
increasing of neutron number up to $N \sim 196$ on going to the 
$Z\sim 96, N\sim 196$ region of octupole deformation and 
especially pronounced with an additional increase of neutron number up to 
$Z\sim 120$. In the $Z\sim 120, N\sim 190$ region, there is a substantial
discrepancies in model predictions. Note that in this region of nuclear chart
the state-of-the-art theories disagree even in the prediction of large 
spherical shell gaps and thus of the properties of superheavy nuclei 
\cite{BRRMG.99,SP.07,AANR.15}.

\section{The impact of pairing strength changes}
\label{pairing}

   The extrapolations beyond the known region of nuclei are 
associated with theoretical uncertainties. The systematic uncertainties 
related to the form of the functional were quantified in Sec.\ 
\ref{uncertainties}; note that they are related to the particle-hole 
channel of the DFTs. In addition, there are the uncertainties in the 
particle-particle (pairing) channel; they are expected to become 
especially large in the vicinity of the two-neutron  drip line (see Refs.\ 
\cite{PMSV.13,AARR.15}). The study of $^{218-234}$Th isotopes in Sect.\ V 
of Ref.\ \cite{AAR.16} showed that in general pairing counteracts the 
shell effects. As a result, the strongest trend towards octupole 
deformation is seen in the systems with no pairing, while the increase 
of pairing suppresses it. The modification of the pairing strength 
may also lead to the changes in the topology of potential energy 
surfaces.

\begin{figure*}
  \includegraphics[angle=0,width=5.9cm]{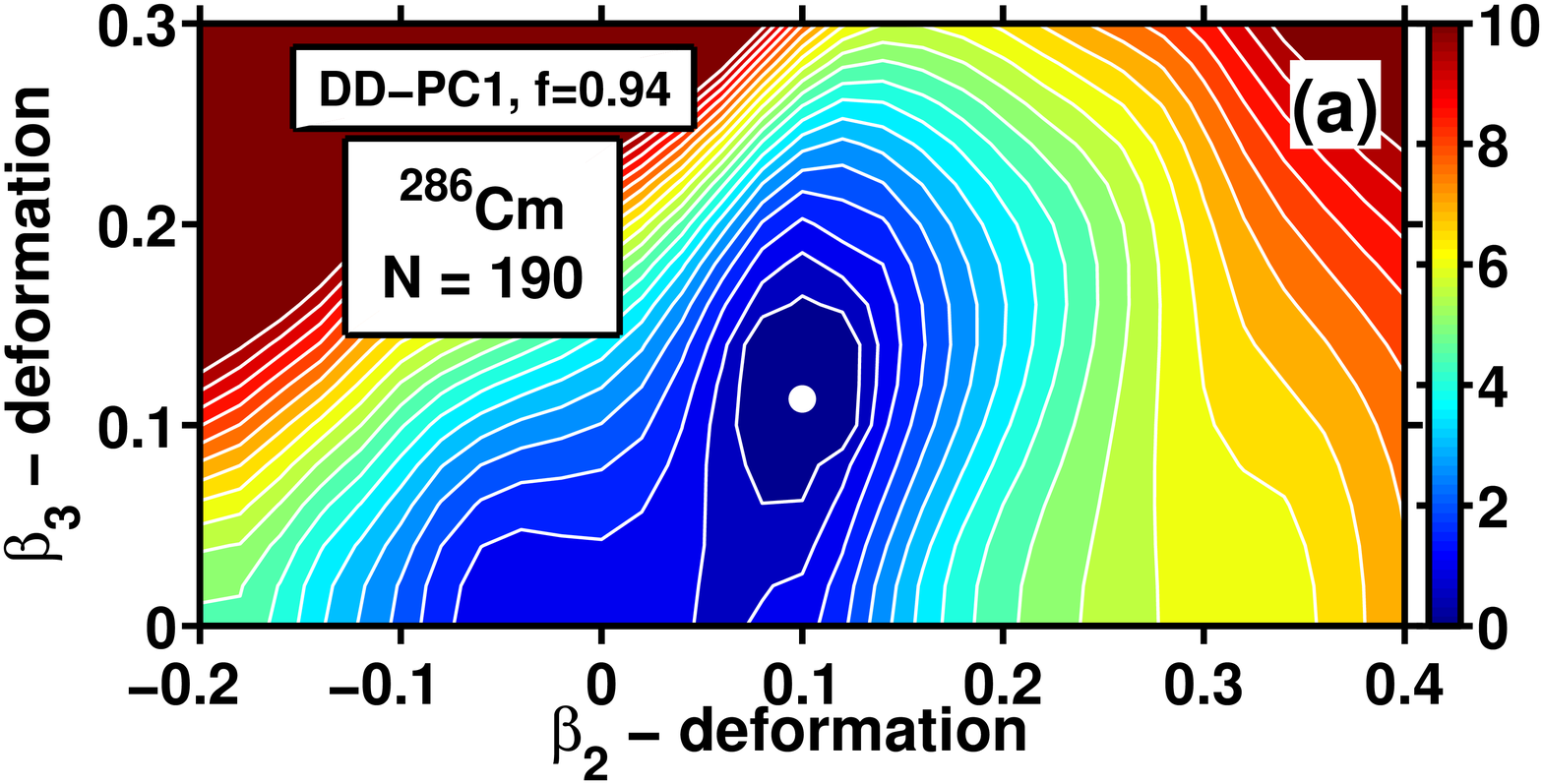}
  \includegraphics[angle=0,width=5.9cm]{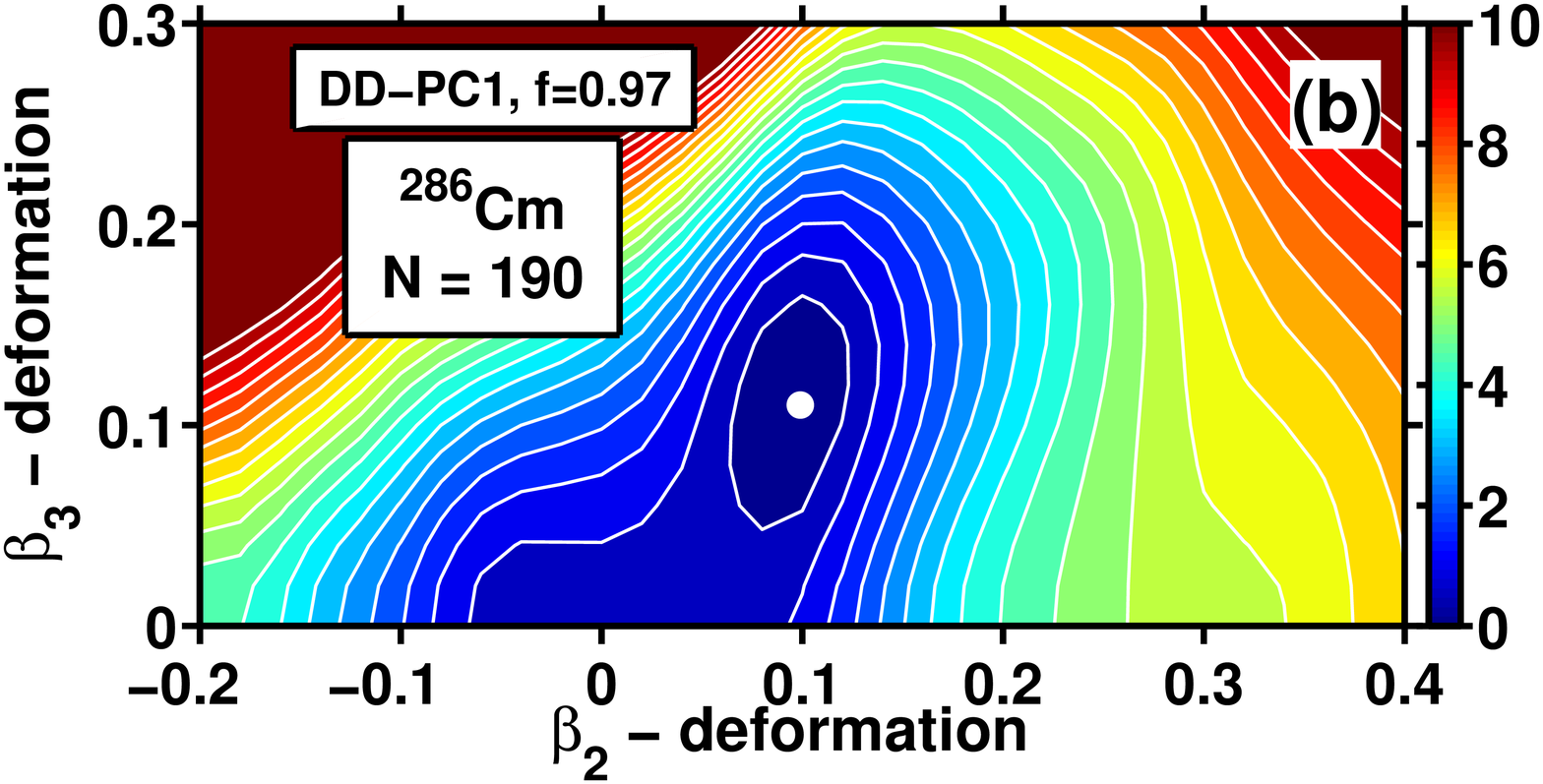}
  \includegraphics[angle=0,width=5.9cm]{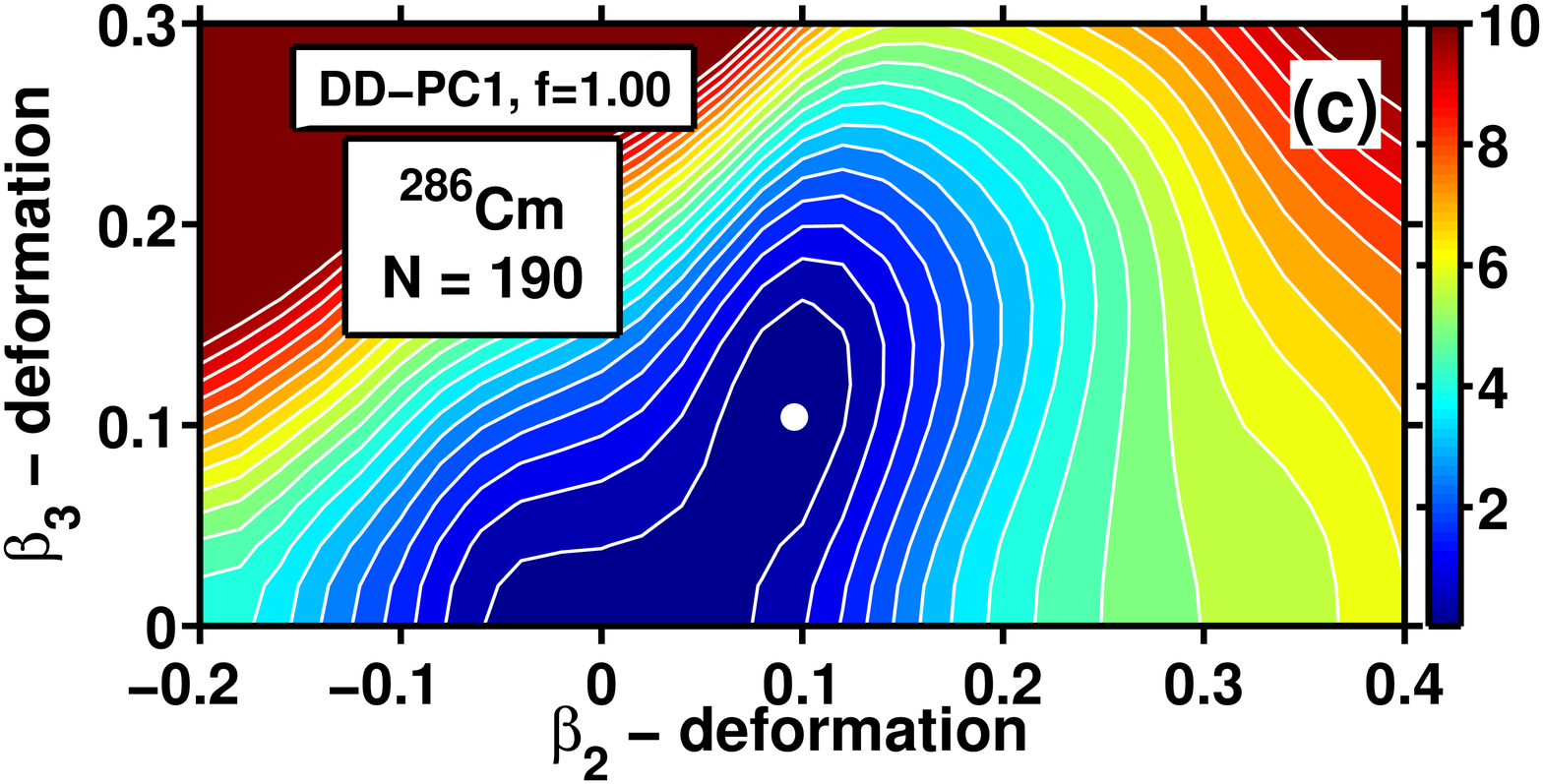}
  \includegraphics[angle=0,width=5.9cm]{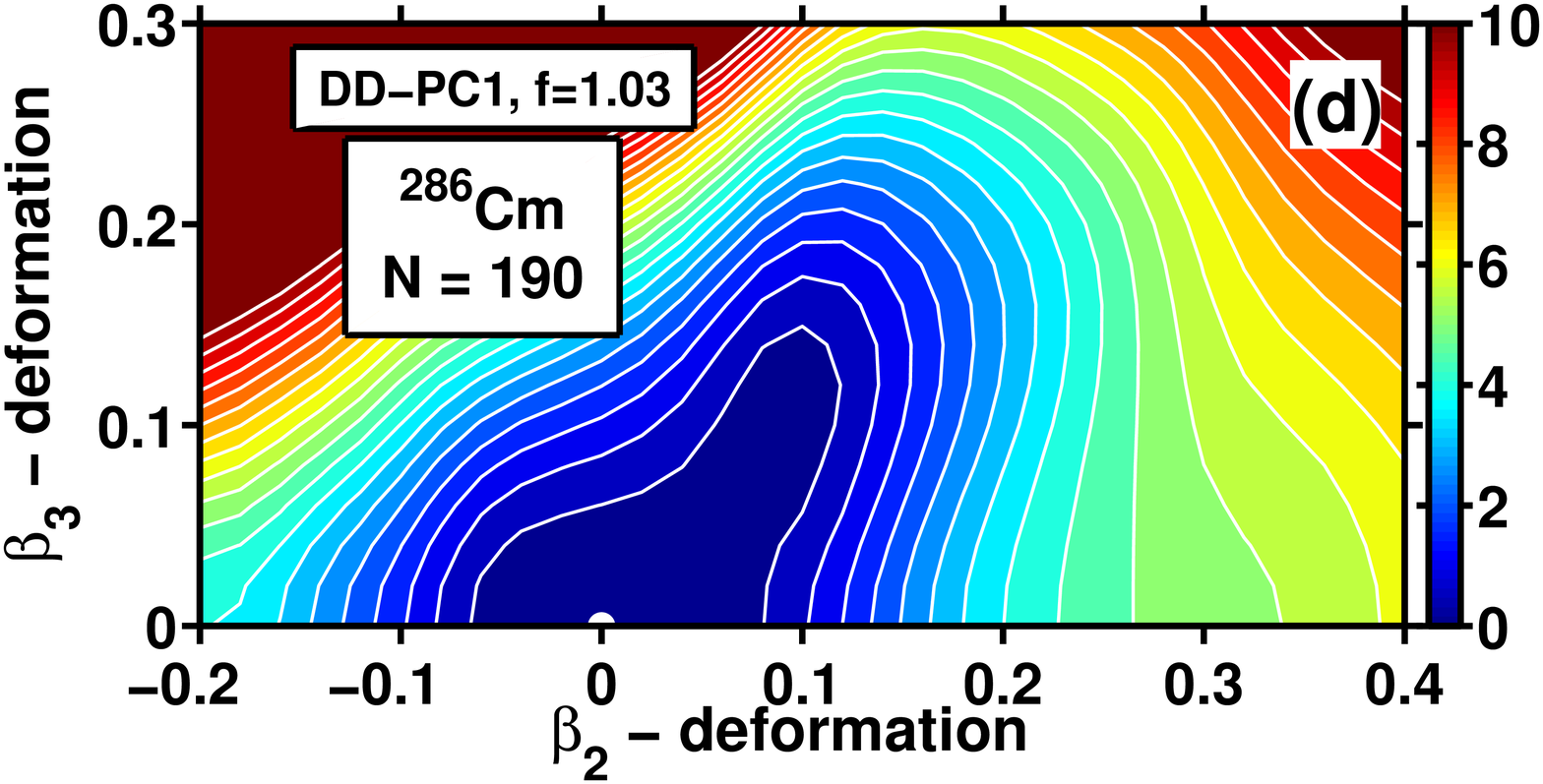}
  \includegraphics[angle=0,width=5.9cm]{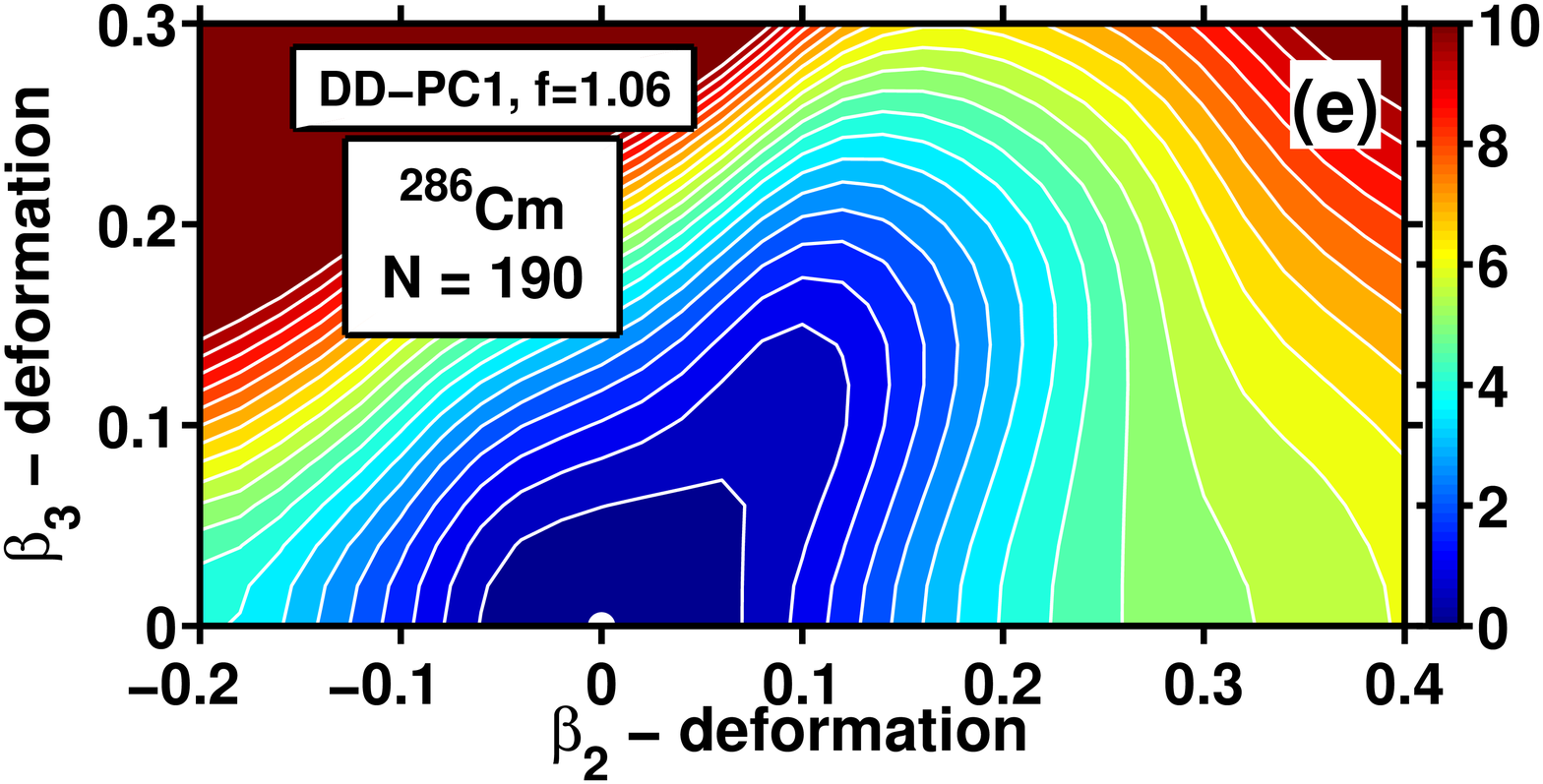}
  \caption{(Color online) Potential energy surfaces of  $^{286}$Cm in 
          the $(\beta_2,\beta_3)$ plane calculated with the CEDF 
          DD-PC1 for different values of scaling factor $f$ of the 
          pairing strength.  White
          circle indicates the global minimum. Equipotential
          lines are shown in steps of 0.5 MeV.}
\label{Cm286_pairing}
\end{figure*}

\begin{figure*}
  \includegraphics[angle=0,width=5.9cm]{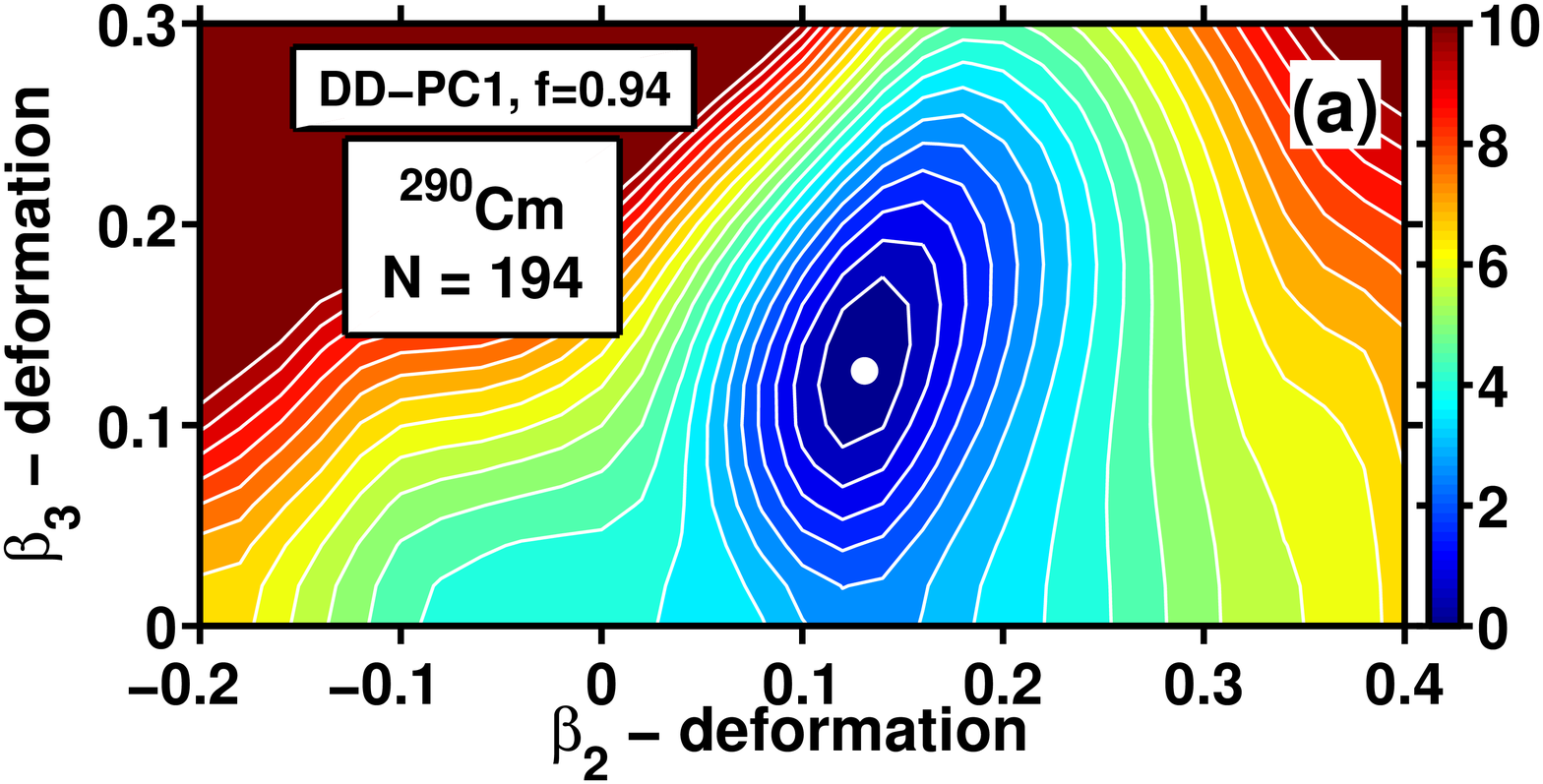}
  \includegraphics[angle=0,width=5.9cm]{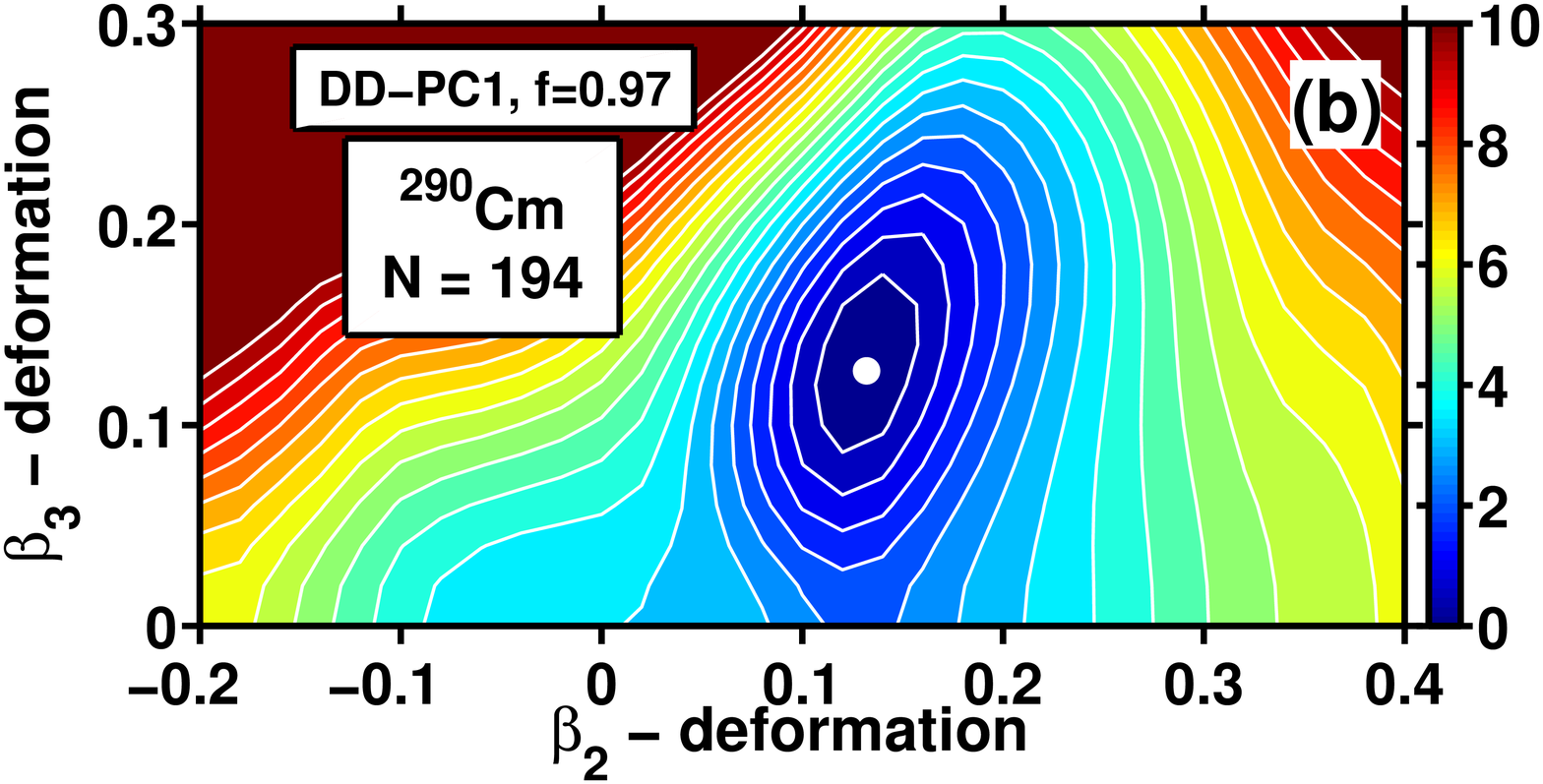}
  \includegraphics[angle=0,width=5.9cm]{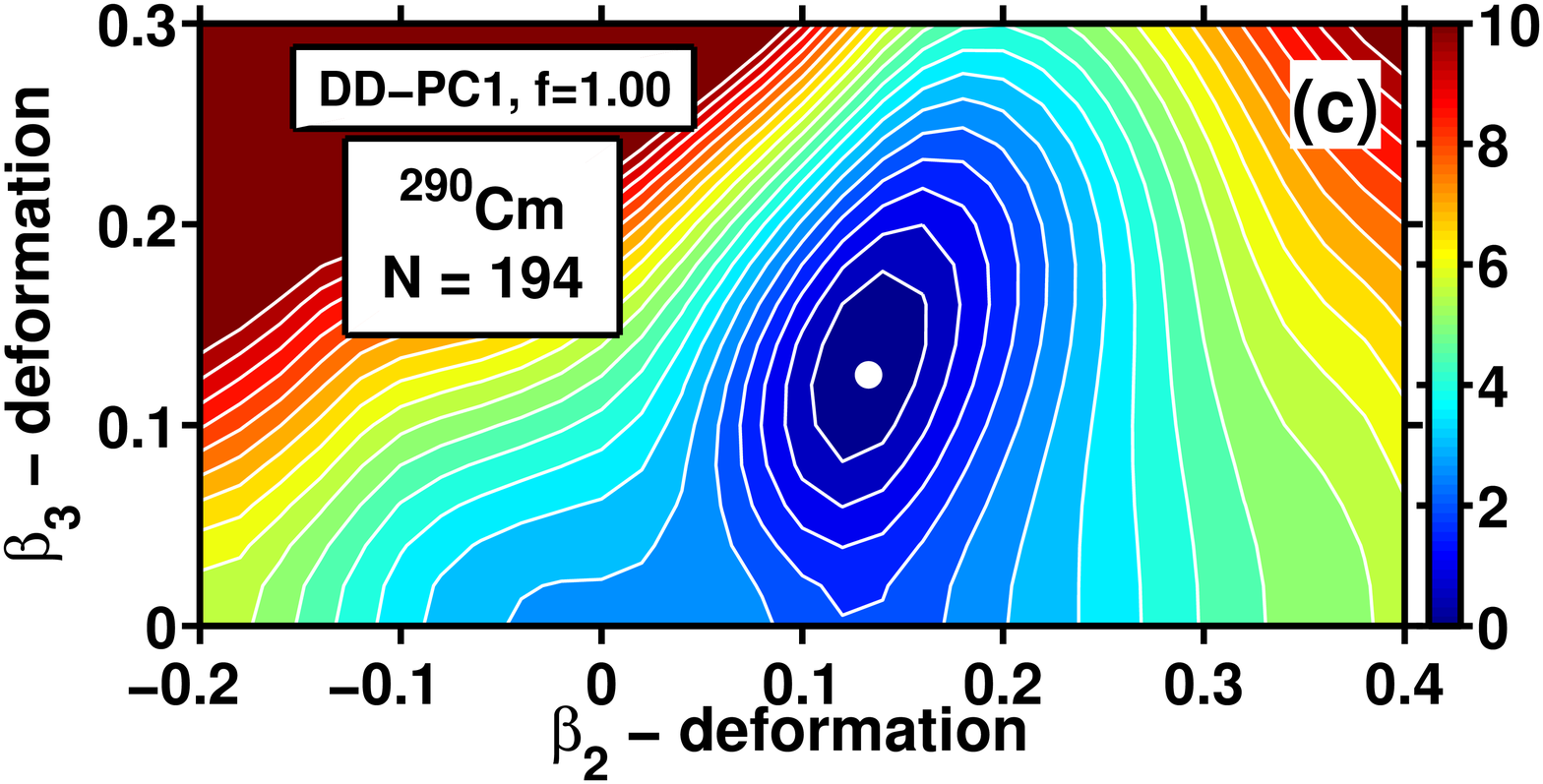}
  \includegraphics[angle=0,width=5.9cm]{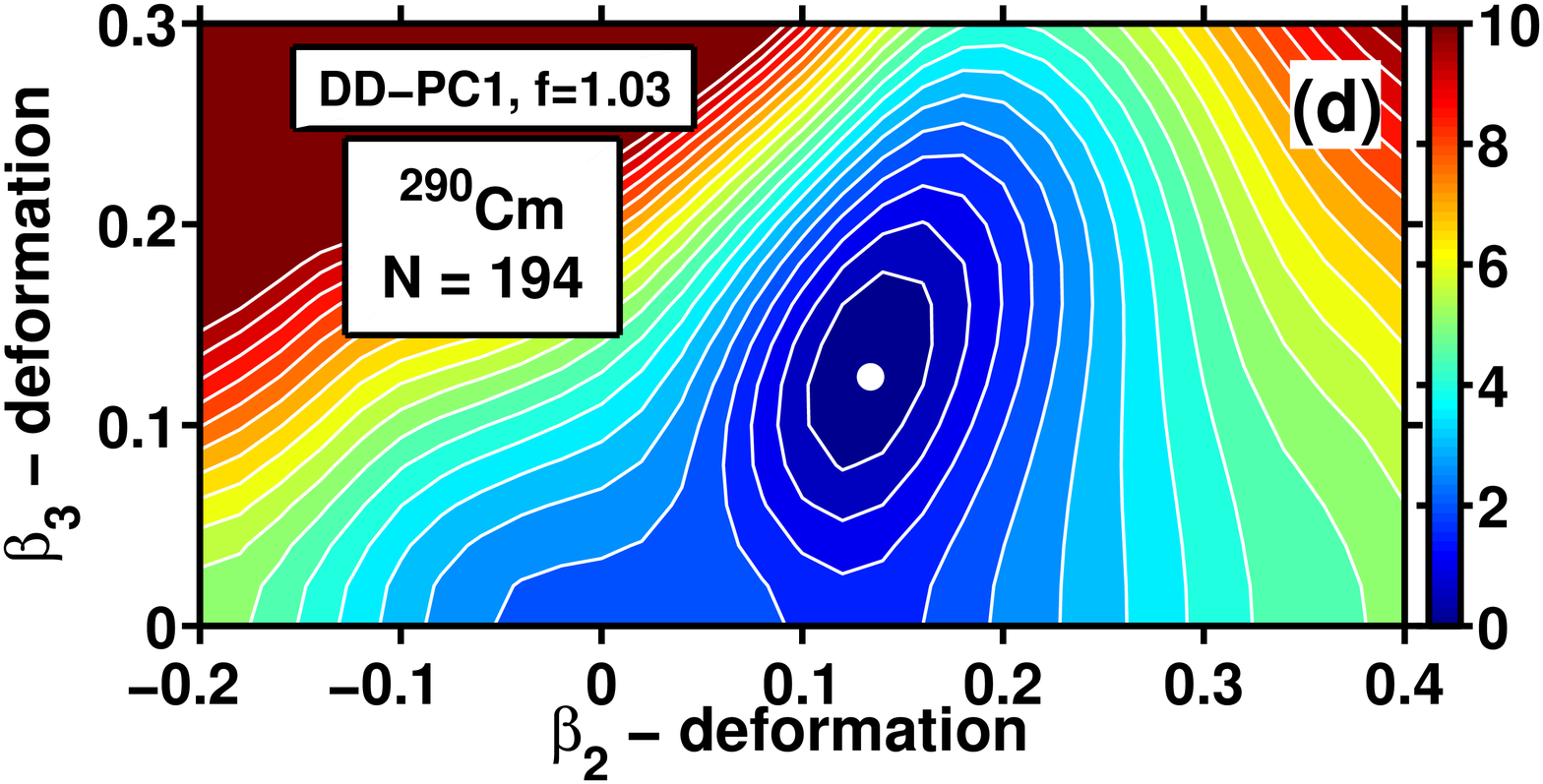}
  \includegraphics[angle=0,width=5.9cm]{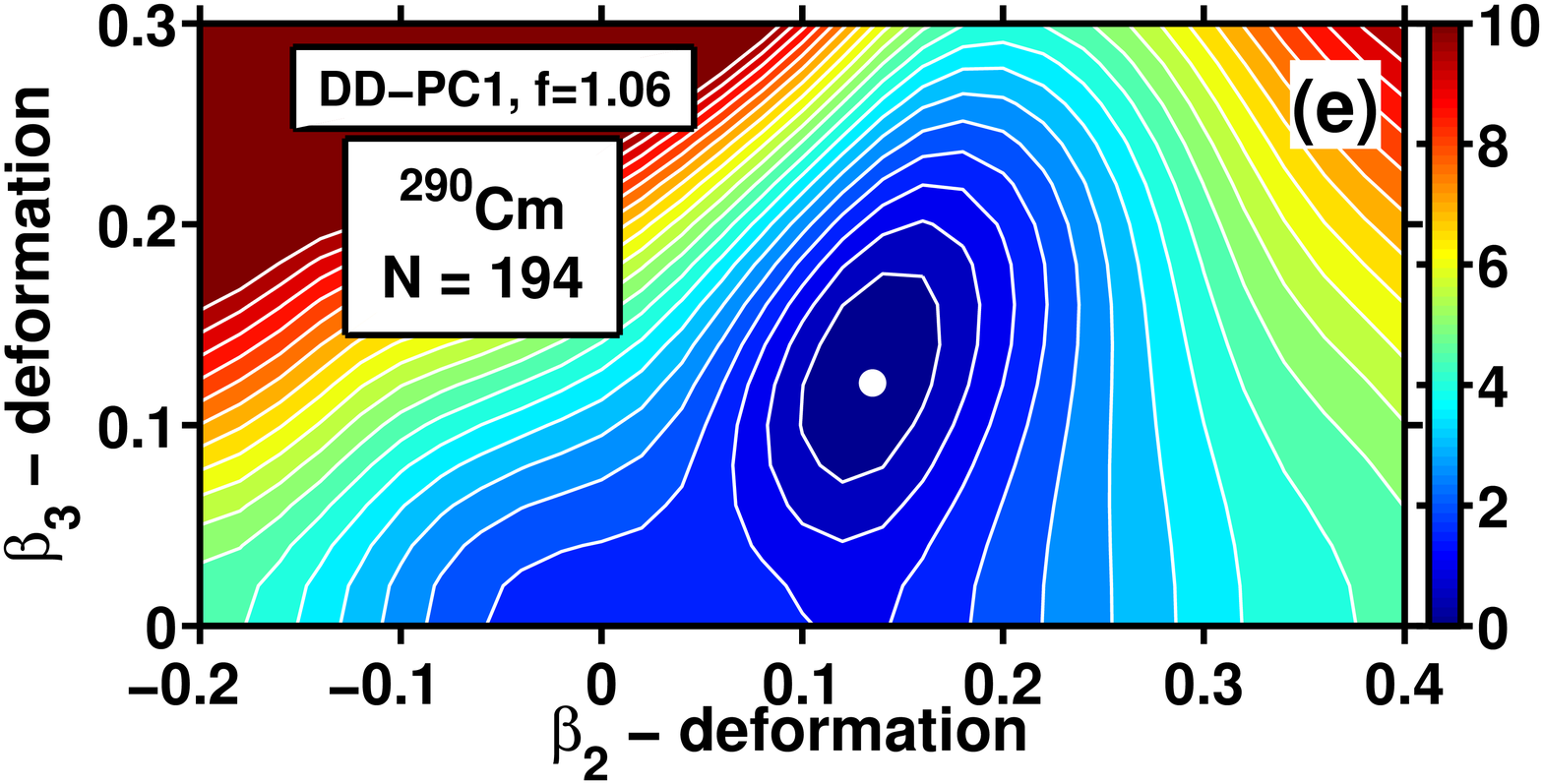}
 \caption{(Color online)  The same as Fig.\ \ref{Cm286_pairing},
           but for $^{290}$Cm.}
\label{Cm290_pairing}
\end{figure*}

 As illustrated in Figs.\ \ref{Cm286_pairing}  and \ref{Cm290_pairing} 
these features are also present in neutron-rich actinides. The $^{286}$Cm 
and $^{290}$Cm nuclei are used here as the examples and the scaling factor
$f$ of the pairing strength is varied in indicated range. This is a factor by 
which the matrix elements of Eq.\ (\protect\ref{TMR}) are multiplied.
Based on previous studies of the pairing in the CDFT framework in Refs.\ 
\cite{AO.13,AARR.15} the variations of the scaling factor in the range of 
$\pm 3\%$ with respect of $f=1.0$ should be considered as most reasonable, 
but still larger variations could not be excluded. The $^{286}$Cm nucleus, 
located at the borderline of the octupole deformed region (see Fig.\ 
\ref{fig-global-CDFT}c), is characterized by PES which is extremely soft 
in octupole direction (Fig.\ \ref{Cm286_pairing}c). The $^{290}$Cm is located 
at the center of the island of octupole deformation (Fig.\ \ref{fig-global-CDFT}c) 
and is characterized by deep octupole minimum with large $|\Delta E^{oct}| \sim 2.0$ 
MeV (see Fig.\ \ref{truncation}d). The impacts of the scaling factor $f$ changes 
on the gain in binding due to octupole deformation and on equilibrium
deformations are summarized in Tables \ref{table-f-impact-binding} and 
\ref{table-f-impact-deformation}, 
respectively. Similar to the results presented in Sec.\ V of
Ref.\ \cite{AAR.16}, the reduction of pairing strength leads to 
more pronounced octupole minimum in both nuclei. On the contrary,
the increase of pairing strength reduces the depth of octupole
minimum in $^{290}$Cm and makes the $^{286}$Cm nucleus spherical. Thus,
one can conclude that weaker (stronger) pairing would make the island 
of octupole deformation broader (narrower) with more (less) pronounced
gains in binding due to octupole deformation in nuclei. The impact of
the modification of the pairing strength on the equilibrium deformation
is small in $^{290}$Cm. Similar situation exists also in $^{286}$Cm for 
$f=0.94-1.00$. However, further increase of $f$ triggers transition
to spherical shape.

\begin{table}[h]
\begin{center}
\caption{The gain in binding $|\Delta E^{oct}|$ (in MeV) due 
to octupole deformation calculated for different values of 
scaling factor $f$ of the pairing.}   
\label{table-f-impact-binding}
\begin{tabular}{|c|c|c|c|c|c|} \hline 
    Nucleus                     &   $f=0.94$  & $f=0.97$ &  $f=1.00$ & $f=1.03$  & $f=1.06$  \\ \hline
    $^{286}$Cm                   &    1.089    & 0.696    &  0.271    &  0.0      &  0.0       \\       
    $^{290}$Cm                   &    2.680    & 2.363    &  1.994    &  1.735    &  1.434      \\ \hline         
\end{tabular}
\end{center}
\end{table}

\begin{table*}[ht]
\begin{center}
\caption{The ($\beta_2, \beta_3$) deformations of the minimum of PES
obtained in the RHB calculations with different values of scaling 
factor $f$.}   
\label{table-f-impact-deformation}
\begin{tabular}{|c|c|c|c|c|c|} \hline 
    Nucleus                     &   $f=0.94$     &  $f=0.97$      &  $f=1.00$      &   $f=1.03$    &  $f=1.06$      \\ \hline
    $^{286}$Cm                   & 0.100, 0.113   &  0.099, 0.110  &  0.095, 0.105  &  0.00, 0.00   &  0.00, 0.00    \\          
    $^{290}$Cm                   & 0.131, 0.127   &  0.132, 0.127  &  0.131, 0.126  & 0.134, 0.124  &  0.135, 0.121  \\ \hline         
\end{tabular}
\end{center}
\end{table*}

\section{Conclusions}
\label{conclusions}

  A systematic search for axial octupole deformation has been performed
in the actinides and superheavy nuclei for proton numbers $Z=88-126$
and neutron numbers from two-proton drip line up to $N=210$ using 
four state-of-the-art covariant energy density functionals. Systematic 
theoretical uncertainties in the description of physical observables 
of octupole deformed nuclei have been estimated. The main results can 
be summarized as follows:

\begin{itemize}
\item
  The present CDFT investigation confirms our earlier predictions 
on the existence of the region of octupole deformation centered
around $Z\sim 96, N\sim 196$ obtained with the DD-PC1 and NL3* 
functionals \cite{AAR.16}. Most of the CEDFs predict the size of 
this region in the $(Z,N)$ plane larger than the one at $Z\sim 92, 
N\sim 136$. On the other hand, 
the impact of octupole deformation on the binding energies of 
the nuclei in these two regions are comparable. Similar region
of octupole deformation is predicted also in Skyrme DFT \cite{ELLMR.12} and 
mic+mac \cite{MNMS.95} calculations. However, it is centered at $Z\sim 100, 
N\sim 190$ in the Skyrme DFT calculations and at $Z\sim 100, N\sim 184$ 
in mic+mac calculations.

\item
  Systematic theoretical uncertainties in the predictions of 
quadrupole ($\beta_2$) and octupole ($\beta_3$) deformations 
as well as the gain in binding due to octupole deformation 
$|\Delta E^{oct}|$ have been quantified within the CDFT 
framework. They are comparable in the $Z\sim 96, N\sim 196$ 
and $Z\sim 92, N\sim 136$ regions of octupole deformation.

\item
  The search for octupole deformation in the ground states 
of even-even superheavy $Z=108-126$ nuclei has been performed in
the CDFT framework for the first time. With exception of two
$Z=108$ (two $Z=108$ and one $Z=110$) octupole deformed nuclei 
in the calculations with CEDF DD-PC1 (DD-ME2), we do not find
octupole deformed shapes in the ground states of these nuclei.
These results are in agreement with the ones obtained in the 
Skyrme DFT but disagree with the ones obtained in Gogny DFT 
and mic+mac calculations. The latter calculations indicate
the presence of large island of octupole deformed $Z>110$
nuclei centered around $N\sim 190$.  These differences in the
location of the islands of octupole deformed nuclei are due to 
the differences in the underlying single-particle structure.

\end{itemize}

\section{Acknowledgements}

  This material is based upon work supported by the U.S. Department
of Energy, Office of Science, Office of Nuclear Physics under Award
No. DE-SC0013037.

\bibliography{references17}
\end{document}